\documentclass[aps,twocolumn,10pt,superscriptaddress,showkeys]{revtex4-1}
\usepackage[colorlinks=true,linkcolor=red,citecolor=blue,urlcolor=blue]{hyperref}
\usepackage{amssymb}
\usepackage{epsfig} 
\usepackage{amsmath} 
\usepackage{graphicx}
\usepackage{color}
\usepackage{float}
\usepackage{xcolor}
\usepackage{amssymb}
\usepackage{enumitem}
\usepackage{multirow}
\usepackage{natbib}
\definecolor{red}{rgb}{0.8,0,0}
\definecolor{violet}{rgb}{0.4,0,0.4}
\definecolor{green}{rgb}{0,0.5,0.0}
\definecolor{navy}{rgb}{0.0,0.0,0.6}
\definecolor{orange}{rgb}{0.8,0.2,0.0}
\usepackage[normalem]{ulem}  % \sout{old text} for strikeout

\newcommand{\apjl}{Astro. Phys. J. Lett.}
\newcommand{\mnras}{Mon. Not. Roy. Astron. Soc.}
\newcommand{\aap}{Astronomy $\&$ Astrophysics}
\newcommand{\nphysa}{Nuc. Phys. A}

\newcommand{\physscr}{Phys. Scr.}
\newcommand{\Natureastro}{Nat. Astron.}
\newcommand{\science}{Science}

\begin{document} 

\title{Hybrid stars are compatible with recent astrophysical observations}

\author{Anil Kumar}
\email{anil.1@iitj.ac.in}
\affiliation{Indian Institute of Technology Jodhpur, Jodhpur 342037, India} 

\author{Vivek Baruah Thapa}
\email{thapa.1@iitj.ac.in}
\affiliation{Indian Institute of Technology Jodhpur, Jodhpur 342037, India} 
\affiliation{National Institute for Physics and Nuclear Engineering (IFIN-HH),
RO-077125 Bucharest, Romania}

\author{Monika Sinha}
\email{ms@iitj.ac.in}
\affiliation{Indian Institute of Technology Jodhpur, Jodhpur 342037, India} 
\date{\today}
\begin{abstract}
{Compact stars (CS) are stellar remnants of massive stars. Inside CSs the density is so high that matter is in subatomic form composed of nucleons. With increase of density of matter towards the centre of the objects other degrees of freedom like hyperons, heavier non-strange baryons, meson condensates may appear. Not only that at higher densities, the nucleons may get decomposed into quarks and form deconfined strange quark matter (SQM). If it is so then CSs may contain SQM in the core surrounded by nucleonic matter forming hybrid stars (HSs). However, the nature and composition of matter inside CSs can only be inferred from the astrophysical observations of these CSs. Recent astrophysical observations in terms of CS mass-radius (M-R) relation and gravitational wave (GW) observation indicate that the matter should be soft in the intermediate density range and stiff enough at higher density range to attain the maximum possible mass above $2~M_\odot$ which is not compatible with pure hadronic equation of states (EOSs). Consequently, we study the HS properties with different models of SQM and find that within vector bag model considering density dependent bag parameter, the model goes well with the astrophysical observations so far.}
\end{abstract}
%\keywords{neutron stars; equation of state; quark matter; gravitational waves}
% The mixed phase of normal baryons and (anti)kaon condensation doesn't exist in neutron star matter for the density-dependent coupling models.
\maketitle

\section{Introduction}\label{intro}

Stellar remnant of type-II supernova explosions can be either a black hole or %neutron star (NS) 
            CS depending on the mass of progenitor star. We can infer their properties by observing the radiation from them in the form of electromagnetic as well as GW radiation recently detectable. The astrophysical observations from the CSs are very interesting and useful to probe the matter properties at extremely high density. The CSs possess mass $\sim 1-2~M_\odot$ with very small radius $\sim 10-13$ km. Consequently, the average density of matter inside these CSs are of the order of $10^{14}$ g/cm$^3$. Naturally the density of matter increases from surface to centre to maintain the hydrostatic equilibrium. The matter density inside these CSs becomes few times of nuclear saturation density ($n_0$) near the center \citep{1996cost.book.....G}. Such high matter density scenarios are impossible to reproduce let alone study the same through any of the terrestrial experiments till date. Hence, CSs %are 
            having such environment %that can 
            provide us the platform %information about this 
            to probe this kind of highly dense matter beyond nuclear saturation density. 
            
            In recent years we obtained plenty of information about these %compact objects
            CSs through electromagnetic spectrum and GW observations. Analysis of these observations put constraints on macroscopic properties like mass, radius and tidal deformability. NICER's (Neutron star Interior Composition ExploreR) observations of two compact objects (PSR J0030+0451 and PSR J0740+6620) helped us to put further constraints on composition of this matter. But still exact matter composition of these compact objects is under investigation.
            
            The outer crust of the CSs are composed of ions and electrons. With the increase in density towards centre, electrons are pushed into the nuclei and combing with the protons of the nuclei produce neutrons inside the nuclei. Hence, in the inner crust the matter is composed of neutron-rich nuclei and electrons. With further increase in density, when density reaches of the order of $10^{11}$ g/cm$^3$, neutrons drip out off neutron-rich nuclei and at the base of the inner crust matter is composed of free neutrons with some neutron-rich nuclei and electrons. In the outer core, as density increases further, the nuclei merge together and at around density of order of $10^{14}$ g/cm$^3$ the matter is composed of mainly neutrons with small admixture of protons and electrons. So inside the core of a CS, the matter is mainly asymmetric nuclear matter. Recent discoveries of massive CSs \cite{2013Sci...340..448A,2020NatAs...4...72C,2021ApJ...915L..12F,2021ApJ...918L..28M,2021ApJ...918L..27R} clearly indicate the presence of matter with density a few times of $n_0$ inside the inner core of the CSs. At that much high %At 
            density%few times nuclear saturation density
            , nucleon Fermi energy is high enough  %have sufficient 
            for giving chance to new degrees of freedom like exotic baryon spectrum \citep{glendenning1991reconciliation,2012Astro,2012NuPhA.881...62W,2010PhRvC..81c5803D} as well as deconfined quarks \citep{1996cost.book.....G} to appear in the inner core of the CSs. Even astrophysical observations discard the possibility of existence of CS entirely composed of pure nucleonic matter. If the pure nucleonic matter is modelled consistent with observations of massive CSs, then the matter is too stiff to reproduce the maximum limit of tidal deformability inferred from the GW observations \cite{2021MNRAS.507.2991T}.
            
            Therefore, the appearance of exotic degrees of freedom demands extensive studies in the context of highly dense matter inside the stellar CSs. There are many possibilities in view of the appearance of exotic degrees of freedom at high density regime. As already mentioned, the possibilities include the appearance of heavier strange \cite{1991PhRvL..67.2414G,2013PhRvC..87e5806C,2016EPJA...52...50O, 2018MNRAS.475.4347R,2018arXiv180107084J,2021NuPhA100922171L,2022CoTPh..74a5302L} and non-strange ($\Delta$) baryons \cite{2014PhRvC..90f5809D,2015PhRvC..92a5802C,2018PhLB..783..234L,2019ApJ...874L..22L,2020PhRvD.102d1301S,2020arXiv201000981B, 2021MNRAS.507.2991T}, appearance of meson condensates \cite{2019arXiv190802042M,2020PhRvD.102l3007T,1982ApJ...258..306H, PhysRevD.103.063004} and even the phase transition to deconfined phase of the quarks.  Presence of exotica makes dense matter EOS softer %and
            resulting in less massive CSs \citep{PhysRevC.53.1416,PhysRevC.62.035801}. With some recent models of highly dense matter with hyperons the maximum mass has been achieved near $2.2~M_\odot$, \textcolor{black}{the lower limit ($2.18M_\odot$) of recently observed mass of PSR J$0952-0607$ \cite{2022ApJ...934L..18R} is marginally satisfied \citep{2021MNRAS.507.2991T,2022EPJA...58...96C}}. The same problem comes with matter modelled with (anti)kaon condensates. In this work we study the CSs considering the appearance of deconfined quark matter inside the inner core of the stars.
            %If core region of matter is composed of SQM and enclosed by nuclear matter crust then the entity is known as hybrid star (HS) \citep{2002PhRvC..66b5802B,2010JPhG...37b5201B,2011PAN....74.1502Y,2012PhRvD..85k4017S,2016MNRAS.463..571F,2019ApJ...877..139G,Nandi_2018}. 
            
            According to Bodmer-Witten conjecture matter composed of up (u), down (d) and strange (s) quarks or strange quark matter (SQM) can be more stable than nuclear matter at high density %under certain circumstances 
            \citep{PhysRevD.4.1601,PhysRevD.30.272}. However, as the strong inter-quark interaction is not still well understood, we have to rely on some phenomenological model of SQM at high density regimes. Then we can test the models with the available astrophysical observations as astrophysical objects are only suitable environment to contain such kind of matter.
            The SQM is well described by the most used phenomenological MIT bag model \cite{PhysRevD.9.3471} in which the hadrons are considered as the bubble of free quarks confined within a bag. The quark interaction has been included in this model as perturbative correction with non-zero strong coupling constant $\alpha_c$ leading to modified bag model. However, this modification is by some \emph{ad-hoc} way to reproduce the some lattice QCD result. Other way to include the strong interaction between the quarks is by introducing vector interaction between them via coupling to a vector field which is popularly known as vector bag (vBAG) model \citep{2019ApJ...877..139G,1997NuPhA.615..441F,franzon2016effects,2021PhyS...96f5303L}. HS models have already been constructed for SQM with MIT bag models and Nambu-Jona-Lasinio (NJL)-type models \citep{2012Astro,2021JPhG...48j5201S,2005ApJ...629..969A,2020arXiv200914274P,2021NuPhA100922171L,2014PhRvC..89a5806O,2022arXiv220800466A}. However, recent astrophysical observations of massive stars demand SQM should have repulsive vector interactions \citep{salinas2019strange,2021PhyS...96f5303L,2021PhRvD.103j3010L}. In this work we consider covariant density functional (CDF) model with density-dependent DD-MEX coupling parameterizations \cite{2020PhLB..80035065T,PhysRevC.105.015802} for hadronic matter and vector bag (vBAG) model for SQM \citep{2016MNRAS.463..571F}. Phase transition from hadronic matter to quark matter can be either smooth (Gibbs construction) or sharp (Maxwell construction). In case of Maxwell construction (MC) the surface tension at the interface is higher than Gibbs construction (GC) \citep{2010JPhG...37b5201B,2017PhRvC..96b5802W,2003NuPhA.723..291V,2007PhRvD..76l3015M}. Vector coupling causes high surface tension at the interface, so we need local charge neutrality that can be possible through MC \citep{2020PhRvD.102b3031X}.
            
            %Recent estimates of mass and radius of several pulsars and tidal deformability of CSs from GW observations are useful to constrain the dense matter model as already mentioned.
            Recently obtained lower limit of maximum attainable mass by the CSs is $M=2.35\pm0.17 M_\odot$ from the observation of PSR J$0952-0607$ \cite{2022ApJ...934L..18R}. M-R constraints obtained through analysis of X-ray data from NICER of PSR J0030+451 \cite{2019ApJ2,2019ApJ4} and PSR J0740+6620 \cite{2021ApJ...918L..28M,2021ApJ...918L..27R} help us to further understand properties of dense matter. Raajimakers et. al. \cite{2021ApJ...918L..29R} evaluated radius range $11.39-13.09$ km of $1.4M_\odot$ CS with $95\%$ credibility from this NICER data. Binary system of compact objects generate GWs during merging due to disturbance in their nearby space-time. These GWs have enough high amplitude that can be detected by our modern GW detectors LIGO and Virgo interferometer. 
            Recent GW170817 and GW190425 signals appears to be binary CS merger events \citep{LIGO_Virgo2017c,2020ApJ1}.
            From data analysis of GW170817 combined tidal deformability parameter ($\tilde{\Lambda}$) found to be less than $900$ \citep{LIGO_Virgo2017c}. Interpreting electromagnetic spectrum of GW170817 data with kilonova models lower limit of $\tilde{\Lambda}\leq400$ is deduced \citep{2018ApJ...852L..29R}. Reanalysis of GW170817 data using PhenomPNRT model this parameter is restricted in the range $110\leq{\tilde{\Lambda}}\leq720$ \citep{abbott2019gwtc}. Tidal deformability parameter ($\Lambda$) of $1.4$ solar mass CS is estimated to be in range $70\leq{\Lambda_{1.4}}\leq580$ with $90\%$ credibility \citep{abbott2018gw170817}. Similarly, for low spin prior systems upper bound on parameter $\tilde{\Lambda}$ is found to be less than $600$ for GW190425 event \citep{2020ApJ1}. Additionally, the analysis of GW190814 signal indicates binary coalescence of a black hole and a CS of mass in range $2.5-2.67 M_\odot$. This secondary component mass range lies in mass gap region (in the region where it can be either a BH or CS). As mentioned earlier, the mass of PSR J0952-0607 is in range $M=2.35\pm0.17 M_\odot$ that is the heaviest pulsar observed till date 
            \citep{2022ApJ...934L..18R}. 
            %claimed 
            Upper limit of this mass range emphasis us to think that secondary component of GW190814 could be a CS. However, as already mentioned very stiff matter with only nucleonic component is too stiff to reproduce the upper limit of $\tilde{\Lambda}$ and matter with hyperonic and bosonic components are too soft to attain the observed mass of the CS. Therefore, we study the possibility of existence of HS compatible with all recent astrophysical observations. 
            
            \textcolor{black}{The paper is organized as follows.
In Sec.-\ref{formalism}, we briefly describe the CDF, vBAG model formalisms as well as the phase transition from hadronic to quark matter.
The implications of de-confined quark matter possibility in CSs are shown and discussed in Sec.-\ref{results}.
Sec.-\ref{conclusions} provides the summary and conclusions of this work.}

\textcolor{black}{\textit{Conventions}: We implement the natural units $G=\hbar=c=1$ throughout the work.}

\section{The matter model}\label{formalism}

The matter at lower density end is nuclear matter and after certain density is SQM with electrons. As mentioned in the sec-\ref{intro}, for nuclear matter section we consider the DD-MEX parametrizations within the CDF model and for SQM we consider vBAG model.

\subsection{CDF model for hadronic matter} 
In lower density region near surface of HS the matter constituents are proton, neutron and electrons. The interaction between these nucleons is  mediated via isoscalar-scalar $\sigma$, isoscalar-vector $\omega$ and isovector-vector $\rho$ mesons. The total Lagrangian density of hadronic matter is given as \cite{1996cost.book.....G}
\begin{equation} \label{eqn:001}
    \begin{aligned}
\mathcal{L}_H & = \sum_{N} \bar{\psi}_N(i\gamma_{\mu} D^{\mu} - m^{*}_N) \psi_N + \frac{1}{2}(\partial_{\mu}\sigma\partial^{\mu}\sigma - m_{\sigma}^2 \sigma^2) \\
 & - \frac{1}{4}\omega_{\mu\nu}\omega^{\mu\nu} + \frac{1}{2}m_{\omega}^2\omega_{\mu}\omega^{\mu} - \frac{1}{4}\boldsymbol{\rho}_{\mu\nu} \cdot \boldsymbol{\rho}^{\mu\nu} + \frac{1}{2}m_{\rho}^2\boldsymbol{\rho}_{\mu} \cdot \boldsymbol{\rho}^{\mu}.
    \end{aligned}
\end{equation}
with the covariant derivative given by $D_{\mu} = \partial_\mu + ig_{\omega N} \omega_\mu + ig_{\rho N} \boldsymbol{\tau}_{N3} \cdot \boldsymbol{\rho}_{\mu}$ with $N$ denoting the nucleons.
Eq. \eqref{eqn:001} provides the minimal Lagrangian as it does not take into account the tensor couplings of vector meson to baryons (appears in Hartree-Fock theories \cite{2006PhLB..640..150L, PhysRevC.92.014302}). The anti-symmetric field tensors corresponding to vector meson fields are given by $\omega_{\mu \nu} = \partial_{\mu}\omega_{\nu} - \partial_{\nu}\omega_{\mu}$ and $\boldsymbol{\rho}_{\mu \nu} = \partial_{\mu} \boldsymbol{\rho}_{\nu} - \partial_{\nu}\boldsymbol{\rho}_{\mu}$.
The Lagrangian density for the leptonic part is given by $\mathcal{L}_l=\sum_{l} \bar{\psi}_l (i\gamma_{\mu} \partial^{\mu} - m_l)\psi_l$ with $m_l$ denoting the mass of leptons ($e^-$).
For the isoscalar meson-nucleon couplings, they are defined as
\begin{equation}\label{eqn.dd_isoscalar}
g_{i N}(n)= g_{i N}(n_{0}) f_i(x) \quad \quad \text{for }i=\sigma,\omega
\end{equation}
where, the function is given by
\begin{equation}\label{eqn.func}
f_i(x)= a_i \frac{1+b_i (x+d_i)^2}{1+c_i (x +d_i)^2}
\end{equation}
where $x=n/n_0$ and $a_i$, $b_i$, $c_i$, $d_i$ are parameters which describe the density-dependent nature of saturation properties.
The isovector-vector $\rho$-meson coupling is given by $g_{\rho N}(n)= g_{\rho N}(n_{0}) e^{-a_{\rho}(x-1)}$.

And to maintain the thermodynamic consistency in case of density-dependent coupling model (such as DD-MEX considered in this work) the rearrangement term $\Sigma^{r}$ is introduced which is given by \cite{2001PhRvC..64b5804H}
\begin{equation}
\begin{aligned}
\Sigma^{r} & = \sum_{N} \left[ \frac{\partial g_{\omega N}}{\partial n}\omega_{0}n_{N} - \frac{\partial g_{\sigma N}}{\partial n} \sigma n_{N}^s + \frac{\partial g_{\rho N}}{\partial n} \rho_{03} \boldsymbol{\tau}_{N3} n_{N} \right].
\end{aligned}
\end{equation}
Note that this term explicitly contributes to the matter pressure only.
\textcolor{black}{The rearrangement term enters through the baryonic chemical potential to solely contribute to the matter pressure term.}
Now in order to describe the dense matter, the baryonic and electric charge conservation should be taken into account with ultimately evaluating the baryonic energy density as,
\begin{widetext}
\begin{eqnarray}
\begin{aligned}
\varepsilon_b & = \frac{1}{2}m_{\sigma}^2 \sigma^{2} + \frac{1}{2} m_{\omega}^2 \omega_{0}^2 + \frac{1}{2}m_{\rho}^2 \rho_{03}^2 + \sum_N \frac{1}{\pi^2} \left[ p_{{F}_N} E^3_{F_N} - \frac{m_{N}^{*2}}{8} \left( p_{{F}_N} E_{F_N} + m_{N}^{*2} \ln \left( \frac{p_{{F}_N} + E_{F_N}}{m_{N}^{*}} \right) \right) \right] \\
	 & + \frac{1}{\pi^2}\sum_l \left[ p_{{F}_l} E^3_{F_l} - \frac{m_{l}^{2}}{8} \left( p_{{F}_l} E_{F_l} + m_{l}^{2} \ln \left( \frac{p_{{F}_l} + E_{F_l}}{m_{l}} \right) \right) \right]
\end{aligned}
\end{eqnarray}
\end{widetext}
with $p_{F_j}$, $E_{F_j}$ denoting the Fermi momentum and Fermi energy of the $j-$th fermion in the system.
With this, the baryonic matter pressure is evaluated from the Gibbs-Duhem relation and given by
\textcolor{black}{\begin{widetext}
\begin{eqnarray}
\begin{aligned}
p_m & = -\frac{1}{2}m_{\sigma}^2 \sigma^{2} + \frac{1}{2} m_{\omega}^2 \omega_{0}^2 + \frac{1}{2}m_{\rho}^2 \rho_{03}^2 + \sum_N \frac{1}{12\pi^2} \left[ p_{{F}_N} E^3_{F_N} - \frac{m_{N}^{*2}}{2} \left( 5 p_{{F}_N} E_{F_N} - 3 m_{N}^{*2} \ln \left( \frac{p_{{F}_N} + E_{F_N}}{m_{N}^{*}} \right) \right) \right] \\
	 & + \frac{1}{12 \pi^2}\sum_l \left[ p_{{F}_l} E^3_{F_l} - \frac{m_{l}^{2}}{2} \left( 5 p_{{F}_l} E_{F_l} - 3 m_{l}^{2} \ln \left( \frac{p_{{F}_l} + E_{F_l}}{m_{l}} \right) \right) \right] + n\Sigma^r
\end{aligned}
\end{eqnarray}
\end{widetext}}

\textcolor{black}{For the outer and inner crust regions, we have implemented the Baym-Pethick-Sutherland \cite{1971ApJ...170..299B} and Negele $\&$ Vautherin \cite{1973NuPhA.207..298N} EOSs respectively which satisfy the nuclear physics data while maintaining thermodynamic consistency in the crust-core transition region \cite{2016PhRvC..94c5804F}.}

\subsection{vBAG model for quark matter}
 Near core, matter is SQM composed of quarks u, d, s with electron (e) as a lepton. vBAG model incorporates quark's interaction in MIT bag model \cite{PhysRevD.9.3471,PhysRevD.30.2379} %with 
 via \textcolor{black}{isoscalar}-vector $V$ field analogous to $\omega$ meson between baryons in CDF formalism \citep{2016MNRAS.463..571F, 2021PhyS...96f5303L,2022MNRAS.513.3788K,2021ApJ...923..250J}. In this model the Lagrangian density of SQM is
\begin{equation}
\begin{aligned}
    {\cal L}_Q & = \sum_{q = u,d,s}\left[\bar{\psi}_q\{\gamma_{\mu}\left(i\partial^\mu - g_{qV}V_\mu\right)-m_q\}\psi_q-B \right] \Theta(\bar{\psi}_q\psi_q) \\ & - \frac{1}{4}{\left(\partial_{\mu}V_{\nu}-\partial_{\nu}V_{\mu}\right)}^2 + \frac{1}{2}{m^2_V}{V_\mu}V^{\mu} +\bar{\psi}_e\left(\gamma_{\mu}i\partial^\mu -m_e\right)\psi_e 
\end{aligned} 
\end{equation}
where $B$ is bag constant and $\Theta$ is heavyside step function (function vanishes outside bag and remains unity inside the bag). $m_V$ is mass of vector meson and $g_{qV}$ is its coupling parameter with quark. The repulsive vector field ($V^\mu$) shifts chemical potential of quark q (u,d,s) as
\begin{equation}
\begin{aligned}
\mu_q = \sqrt{(k_{f_q})^2 + (m_q)^2} + g_{qV}V_0,
\end{aligned}
\label{eqn:aa}
\end{equation}  
where $k_{f_q}$ is the Fermi momentum of quark $q$. %In the ground state of matter only temporal component of $V^\mu$ survives as:
%\begin{equation}
%\begin{aligned}
%m_VV_0 = & \sum_{q = u,d,s} \left(\frac{g_{qV}}{m_V}\right)n_q
%\end{aligned}
%\label{eqn:c}
%\end{equation}
%in this equation on right hand side $n_q$ signifies number density of quark q. The new parameter $G_V=\left({g_{qV}}/{m_V}\right)^2$ assists to simplify previous equations. 
With this model, the energy density of the quark matter is %is given by the following equation for this vBAG model
\begin{equation}
\begin{aligned}
\epsilon & = \frac{3}{\pi^2}\sum_{q}\int_{0}^{k_{f_q}}\left(\sqrt{(k)^2 + (m_q)^2}+ g_{qV}V_0\right)k^2dk \\ & - \frac{1}{2}\left({m_V}{V_0}\right)^2 + B,
\end{aligned}
\end{equation} 
and pressure can be obtained by using Gibbs-Duhem relation as:
\begin{equation}
\begin{aligned}
P = \sum_{q}\mu_qn_q - \epsilon,
\end{aligned}
\end{equation} 
where $n_q$ signifies number density of quark $q$.
%$\mu_q$ indicates chemical potential of individual quark. 

\subsection{Phase transition}
As density increases inside CSs, the distance between adjacent nucleons decreases. \textcolor{black}{Nucleons touch each other t density $\rho\sim(1.6~\text{fm})^{-3}=0.24$ $\text{fm}^{-3}$ assuming these nucleons as spheres of rms radius $0.8$ fm. Considering the merger of nucleons is necessary condition for quarks deconfinement, we consider the threshold density for quark deconfined should be $\rho=0.24$ $\text{fm}^{-3}$ \citep{1996cost.book.....G}.} So, this density of nucleonic matter is the threshold density i.e the lower limit for phase transition (PT) from hadronic matter to quark matter. Neutron chemical potential corresponding to this lower limit of matter density for DD-MEX EOS is $\sim 1039$ MeV which is consistent with the theoretical result for QCD phase transition considering vBAG model \cite{2021PhyS...96f5302L,2022MNRAS.512.5110L}. Also extrapolation to zero temperature of experimental results indicates the same range of nuclear chemical potential \cite{2021PhyS...96f5302L}. %approximately equal to lower limit of chemical potential suggested by refs.- .
In this work we consider %assume 
first order phase transition from hadronic matter to quark matter through MC. MC requires local charge neutrality for both phases. Maxwell conditions for phase transition are :
\begin{equation}
\begin{aligned}
    P_H = P_Q,\\  {\mu_b}^H = {\mu_b}^Q
\end{aligned}
\end{equation}
where $P_H$ and $P_Q$ are hadronic and quark phase pressure, respectively. The baryonic chemical potential $(\mu_b)$ is given by following equilibrium conditions,
\begin{equation}
\begin{aligned}
    \mu_b = \mu_n = 3(\mu_u + \mu_e),\\
    \mu_s = \mu_d = \mu_u + \mu_e
\end{aligned}
\end{equation}
where $\mu_n$ is neutron chemical potential and other chemical potentials are according to notations we mentioned above. 

\subsection{Parameter space}\label{subsec:parameter}

For nucleonic matter EOS, we consider DD-MEX parametrization \cite{2020PhLB..80035065T,PhysRevC.105.015802} as already mentioned. For SQM, we %assume
consider the masses of quarks as $m_u=4$ MeV, $m_d=7$ MeV and $m_s=100$ MeV. In vBAG model the vector interaction for all the three flavors are considered to be same. 

Bag constant $B$ is inward pressure that keeps quarks confined inside the bag. As the density of matter increases, the energy of the bag decreases and the size of nucleons increases \cite{2001NuPhA.695..353L,2003NuPhA.725..127L}. Bag constant $B$ should depend on density of matter \cite{2002PhRvC..66b5802B,2021JPhG...48j5201S,2013Ap.....56..121Y}, multiple type of parametrization is available for the variation of $B$ with density \citep{2004PhRvD..69j3001P}. But the Gaussian parametrization is mostly opted and can be given as:
\begin{equation}
\begin{aligned}
    B(\rho) = B_c + (B_s - B_c)exp\left[-\beta\Large\left(\frac{\rho}{\rho_0}\Large\right)^2\right]
\end{aligned} 
\label{eqn:b}   
\end{equation}
where $B_c$, $B_s$, $\rho$ and $\rho_0$  are bag constant near center, bag constant at surface, baryon number density and nuclear saturation density respectively. Baryon number density $(\rho)$ for quark matter is related to quark number densities through the following relationship,
\begin{equation}
\begin{aligned}
    \rho = \frac{n_u + n_d + n_s}{3}
\end{aligned}    
\end{equation}

%Considering bag constant ($B$) as function of matter density given by equation (\ref{eqn:b}) is more realistic. 
We consider the free parameters, $\beta=0.2$ and ${B_c}^{1/4}=130$ MeV in this work. 
\begin{figure} [h!]
\begin{center}
\includegraphics[width=8.7cm, keepaspectratio]{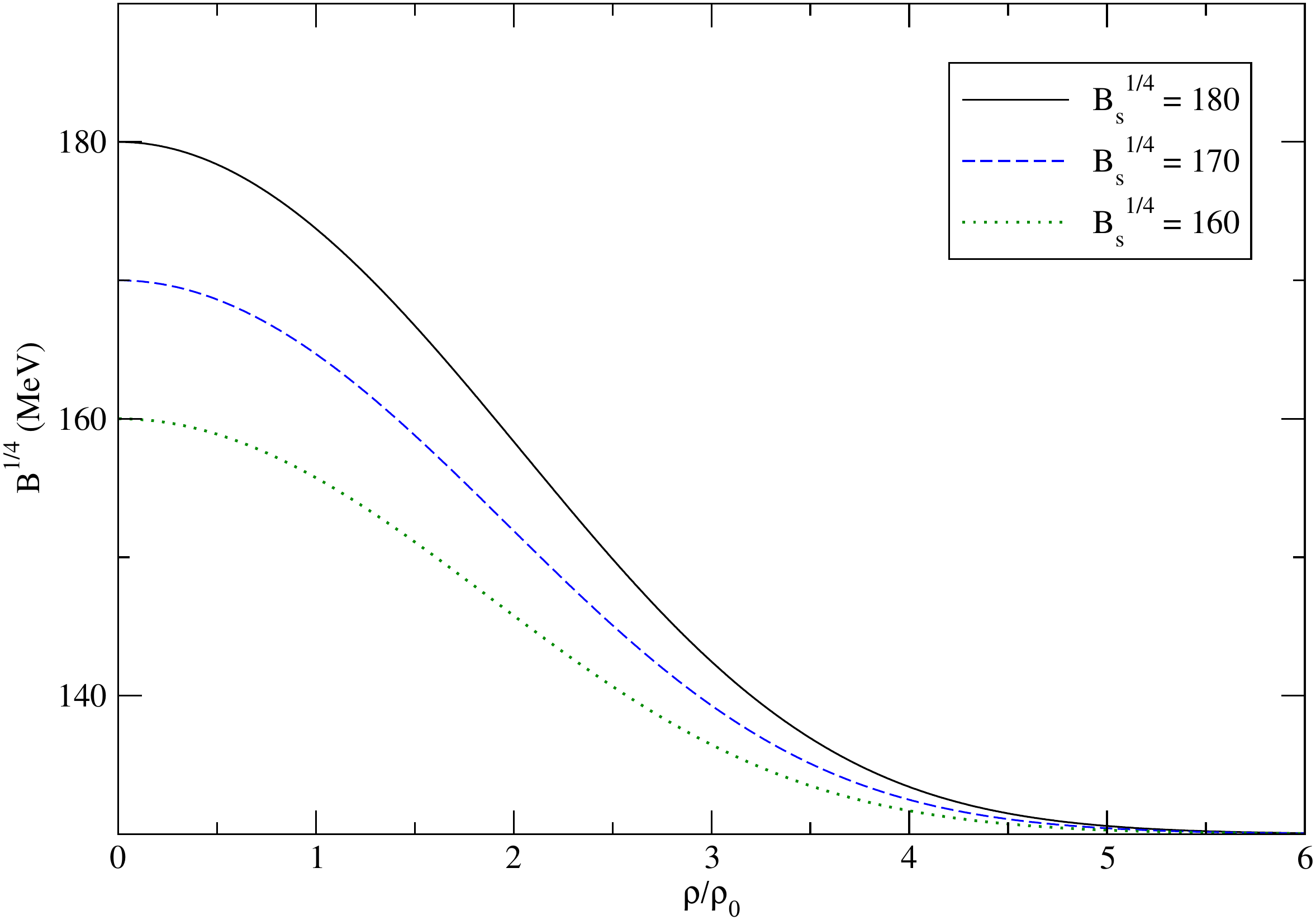}
\caption{Variation of bag parameter $B$ with %as function of 
density for the EOSs we considered with $B_c^{1/4} = 130$ MeV. The solid curve, dashed curve and dotted curve are for $B_s^{1/4} = 180, 170, 160$ MeV respectively.} 
\label{fig-001}
\end{center}
\end{figure} 
 With this choice of parameters $\beta$ and $\gamma$ the variation of $B$ with density is %We assume the $B$ as function of density as 
 shown in Fig. \ref{fig-001}.
\textcolor{black}{Now if $B$ is density dependent then an extra term appears in the expression of chemical potential as 
\begin{equation}
\begin{aligned}
\mu_q = \sqrt{(k_{f_q})^2 + (m_q)^2} + g_{qV}V_0 + \frac{dB(\rho)}{dn_q}.
\end{aligned}
\end{equation} 
}

For stable SQM the range of $G_V=g_{qqV}/m_V$ values are considered between $0-0.3$ fm$^2$ \cite{2021PhyS...96f5303L}. However, in case of HS, stability of quark matter is not required. So we have freedom to choose the value of $g_V$ higher than $0.3$ fm$^2$. For different values of $g_V$, $B_W$ the minimum values of $B$ are shown in the Table \ref{tab:aa} for which the stability of SQM is not assured. 

\begin{figure} [h!]
\begin{center}
\includegraphics[width=8.7cm, keepaspectratio]{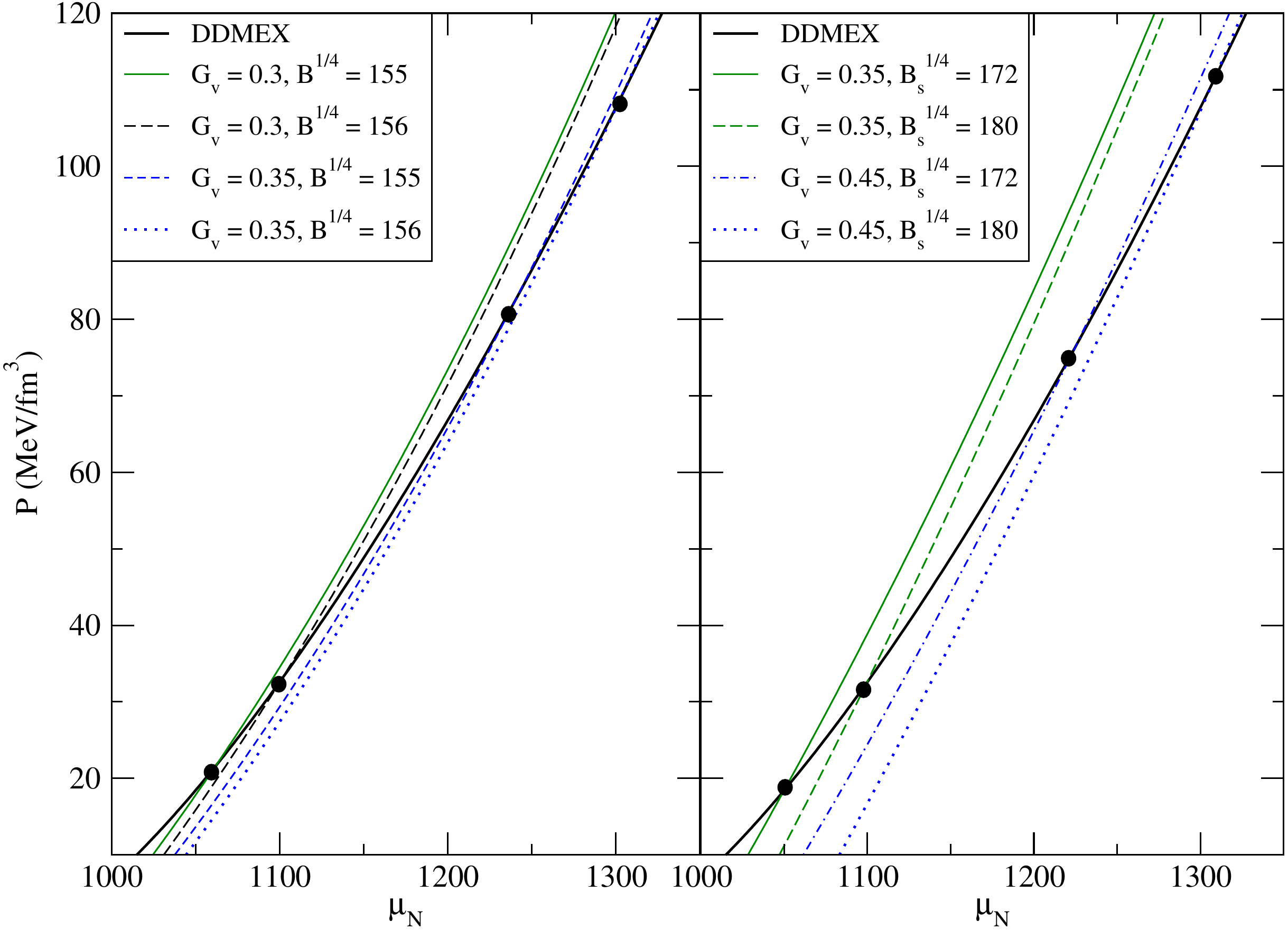}
\caption{Variation of pressure %vs 
with baryon chemical potential  (neutron chemical potential) %of 
in both phases.  Left panel: for density independent $B$ parameters. Right panel: for density dependent $B$ parameters. For both the panels the solid curve is for the nucleonic matter and for SQM matter the parameters values for different curves are as indicated in the figure. The point of PT for each parametrizations of SQM %Intersection points or equilibrium points are 
is denoted by black circles.}
\label{fig-003}
\end{center}
\end{figure} 

\begin{figure*} [t!]
\begin{center}
\includegraphics[width=12cm, keepaspectratio]{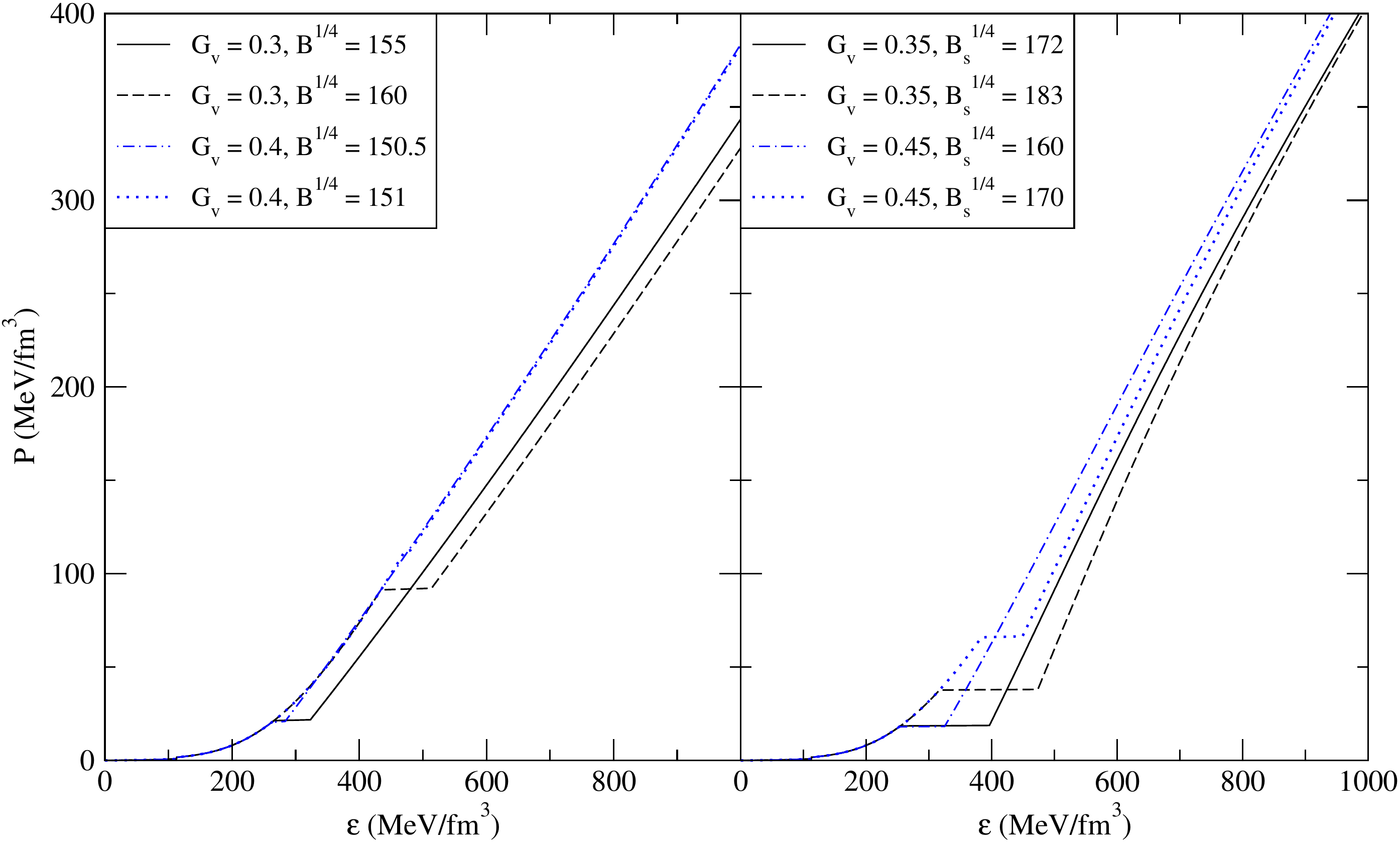}
\caption{Variation of pressure %vs 
with energy density. Left panel: for density independent $B$ parameters. Right panel: for density dependent $B$ parameters. The parameter values for different curves are as indicated in the figure.} 
%relations (EOSs) for all parameters.
 
\label{fig-002}
\end{center}
\end{figure*}

\begin{table}
\begin{center}
\caption{The values of $B_W$ from stability window and $B_{PT}$ from phase transition for density independent (DI) and density dependent (DD) parameterization.}
\begin{tabular}{cccccccccccc}
\hline \hline
      $G_V$ & $B_W^{1/4}$ (DI) & $B_W^{1/4}$ (DD) & $B_{PT}^{1/4}$ (DI) & $B_{PT}^{1/4}$ (DD) \\
    (fm$^2$)& (MeV) & (MeV) & (MeV) & (MeV) \\         
\hline
 0.3 & 146 & 153 & 155 & - \\
 0.35 & 145 & 151 & 152.5 & 172 \\
 0.4 & 144 & 149 & 150.5 & 165 \\
 0.45 & 143 & 147 & - & 160 \\
\hline
\end{tabular}
\label{tab:aa}
\end{center} 
\end{table}

In the Fig. \ref{fig-003} we plot the variation of pressure with baryon chemical potential for nucleonic matter with DD-MEX EOS and for SQM with vBAG model with the different values of $g_V$ and $B$ outside the stability window. The equilibrium points between two phases are shown by dots in Fig. \ref{fig-003}. The pressure point where the baryon chemical potential of nucleonic matter crosses that of the SQM indicates the point of PT. From the figure it is seen that with increase of $G_V$ the baryon chemical potential increases. Again with same value of $G_V$, the baryon chemical potential increases with increase of $B$. So larger the values of $G_V$ and $B$, the later is the PT. Thus the values of $G_V$ is restricted from upper side for PT to occur within the CS. %It is noticed that with increase in $G_V$ the $P-\mu$ curves start overlapping with DDMEX hadronic matter $P-\mu$ curve. So, we get multiple equilibrium points that is unrealistic. Hence} in case of density independent values of $B$ the} left panel we can see that 
Upper value of $G_V$ is restricted to $0.4$ fm$^2$ and in case of density dependent values of $B$, the upper value of $G_V$ is restricted to $0.5~\text{fm}^2$. %In right panel Pressure at equilibrium of both phases is get shifted up with higher values of $B$. 
From the figure it is clear that for a particular value of $G_V$, the chemical potential at transition point decreases with the decrease of $B$. So, this sets a lower limit of $B$ for every value of $G_V$ in the view of that the threshold nuclear chemical potential for the PT from hadronic matter to deconfined SQM is around $1039$ MeV for nucleonic matter with DD-MEX EOS. From this consideration $B_{PT}$, the permissible lowest values of $B$ corresponding to different values of $G_V$ are given in the Table \ref{tab:aa}.  %We get pressure equilibrium points for same values of baryon (neutron) chemical potential in both phases. 

\section{Results}\label{results}

According to the condition of PT from nucleonic matter to SQM the matter model parameters are chosen as discussed in subsection \ref{subsec:parameter}. We have seen that for density independent values of $B$ the maximum admissible values of $G_V$ is $0.45$ fm$^2$ and for density dependent values of $B$ it is $0.5$ fm$^2$. Within these ranges of $G_V$ the minimum values of $B$ for PT to occur is tabulated in Table \ref{tab:aa}. Following these constraints on parameter space we plot the matter EOS in Fig. \ref{fig-002} for different values of $G_V$ and $B$. The matter
%In fig.-, we represents comparison between EOS with constant $B$ and density dependent $B$. EOS 
becomes stiffer with lower values of $B$ and higher values of $G_V$ as expected. According to Maxwell criterion the pressure in both phases remains same at the interface but energy density immediately increases in quark phase. The position of PT point is more sensitive to the values of $B$ compared to the values of $G_V$.%For same values of $G_V$ the PT point for different values of $B$ get shifted. EOSs with density dependent $B$ are stiffer than constant $B$. 
The macroscopic properties of star like mass and radius particularly depends on these equilibrium points.  

\begin{figure*} [t!]
\begin{center}
\includegraphics[width=12cm, keepaspectratio]{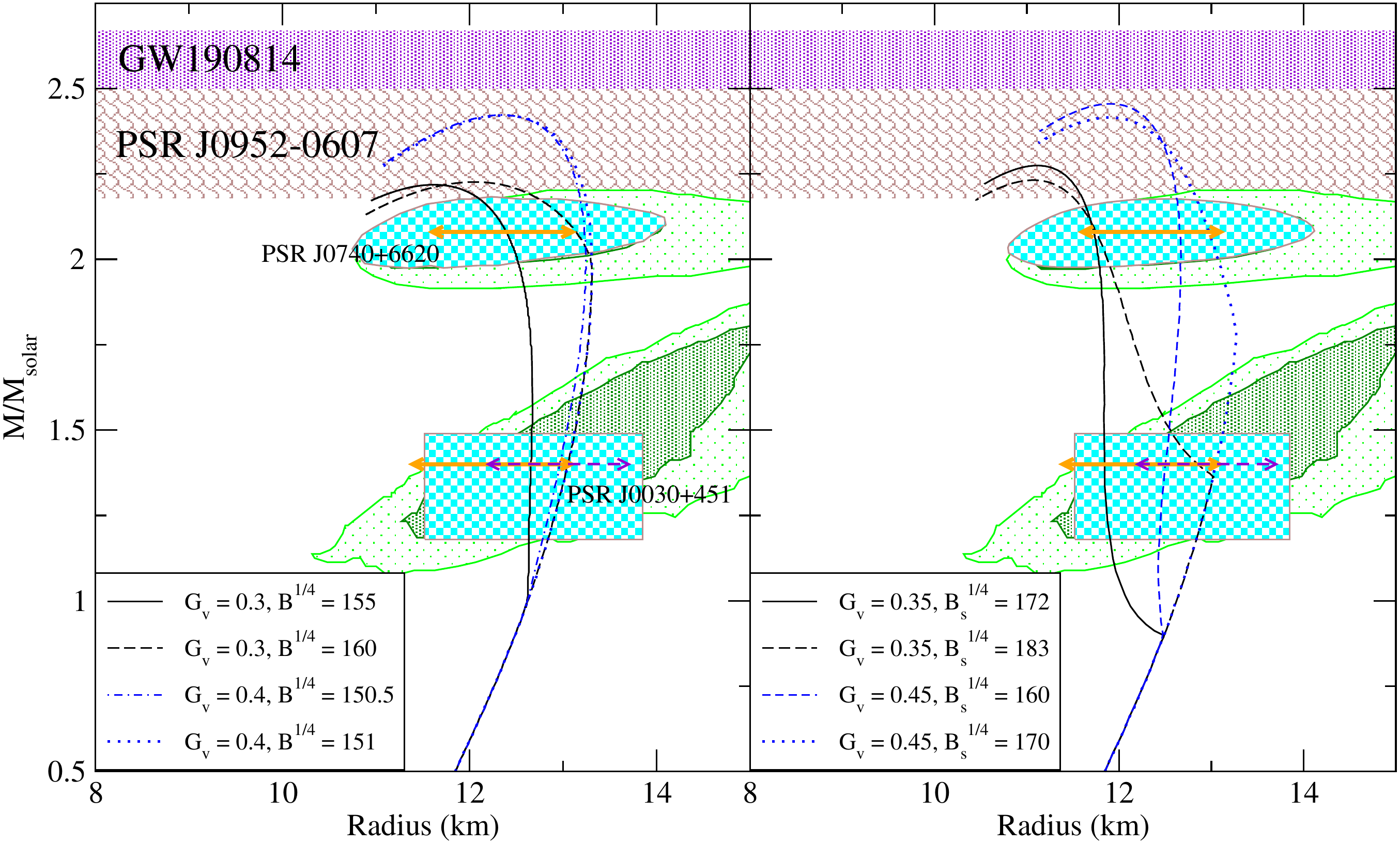}
\caption{Mass-radius relations along with constraints from astrophysical observations for different parametrizations. Left panel: for density independent $B$ parameters. Right panel: for density dependent $B$ parameters. The parameter values for different curves are as indicated in the figure. Contours represent constraints for PSR J0740+6620 \cite{2021ApJ...918L..27R,2021ApJ...918L..28M} and PSR J0030+451 \cite{2019ApJ2,2019ApJ4}. Horizontal straight lines depict radius constraints as we discussed in introduction section. Upper shaded regions are for GW190814 and PSR J0952-0607.}
\label{fig-004}
\end{center}
\end{figure*}   

We construct $M-R$ relations of CS composed of the matter with these EOSs and choices of parameters. %By solving TOV equations for the EOSs, 
We represent those M-R relations along with M-R constraints obtained through astrophysical observations in Fig. \ref{fig-004}. Appearance of SQM at high density softens the matter. Consequently, the stars become more compact compared to stars composed of pure nucleonic matter and earlier appearance of SQM lowers the theoretical attainable maximum mass. It is evident from the figure. For density independent $B$ parameters PT occurs for the stars with masses near $2~M_\odot$ at $B^{1/4} = 160$ MeV if we consider $G_V =0.3$ fm$^2$ and at $B^{1/4} = 151$ MeV if we consider $G_V =0.4$ fm$^2$. This sets the upper limit of $B$ for a specific value of $G_V$. Within this range of $B$ and $G_V$, the theoretical attainable maximum mass comes out to be compatible with the observationally obtained lower limit of CS maximum mass. However, later appearance of SQM does not favour the the smaller radii of CSs as estimated form the NICER's x-ray data analysis for pulsars PSR J$0030+51$ and PSR J$0740+6620$ \cite{2021ApJ...918L..28M,2021ApJ...918L..27R,2019ApJ2,2019ApJ4}. On the other hand if we consider the lower values of $B$ corresponding to each $G_V$ we get PT to occur for stars with masses less than $1.4~M_\odot$. In those cases, though the maximum theoretical attainable mass decreases due to larger portion composed of SQM, still they are within the range of observed lower limit of CS mass. Again we see that in the lower side of $B$, still the matter with higher $G_V = 0.4$ fm$^2$ is too stiff to satisfy the radius constraints obtained through NICER's x-ray data analysis \cite{2021ApJ...918L..28M}. Hence for density independent $B$ case the parameter values are constrained near $G_V = 0.3$ fm$^2$ and $B^{1/4} = 155$ MeV. On the other hand, as for density dependent $B$, the larger valued of $G_V$ is required for PT, the matter may be stiffer compared to density independent $B$ case. Hence, with density dependent $B$, theoretical maximum mass $\sim2.46M_\odot$ can be obtained with $G_V = 0.45$ fm$^2$. However, in this case also if the PT occurs for stars with masses larger than $1.4~M_\odot$, then the the radius constraints for intermediate mass stars obtained through NICER's x-ray data analysis \cite{2021ApJ...918L..29R} can not be satisfied. Therefore, from this observations, we may consider the for density dependent $B$, with $G_V = 0.35$ fm$^2$ the maximum allowed value is $B_s^{1/4} = 183$ MeV and with $G_V = 0.45$ fm$^2$ the maximum allowed value is $B_s^{1/4} \approx 165$ MeV.  %Radius of canonical $1.4$ solar mass star with DDMEX EOS is more than upper limit provided by Raaijmakers et. al. \citep{2021ApJ...918L..29R}. Very early PT from hadronic matter to quark matter reduces radius of $1.4M_\odot$ star. Considering $B$ as density independent in left panel, the higer value of $G_V=0.35$ $\text{fm}^2$ concequences soft matter at higher density. The radius constraints obtained through NICER's x-ray data analysis are marginally satisfied with $B^{1/4}=153$ MeV. Assuming lower value of $G_V=0.3$ $\text{fm}^2$  radius constraints are well satisfied with early PT corresponding to $B^{1/4}=155$. But mass constraint of PSR J0952-0607 is satisfied near its lower bound on mass. On the other hand very high maximum mass ($\sim2.52M_\odot$) is obtained through considering density dependent $B$ in right panel. Multiple set of parameters like ($G_V=0.45$ $\text{fm}^2$,$B_s^{1/4}=165$ MeV), ($G_V=0.45$ $\text{fm}^2$,$B_s^{1/4}=170$ MeV),($G_V=0.48$ $\text{fm}^2$, $B_s^{1/4}=162$ MeV) and ($G_V=0.5$ $\text{fm}^2$,$B_s^{1/4}=160$ MeV) are satisfying the M-R constraints. The larger repulsive vector inetractions with $G_V$ produces soft matter at higher densities and density dependent $B$ produces stiffer EOS as compared to left panel. The only set of parameter ($G_V=0.5$ $\text{fm}^2$,$B_s^{1/4}=160$ MeV) satifies M-R constraints of GW190814 secondary component. 

\begin{figure} [h!]
\begin{center}
\includegraphics[width=8.7cm, keepaspectratio]{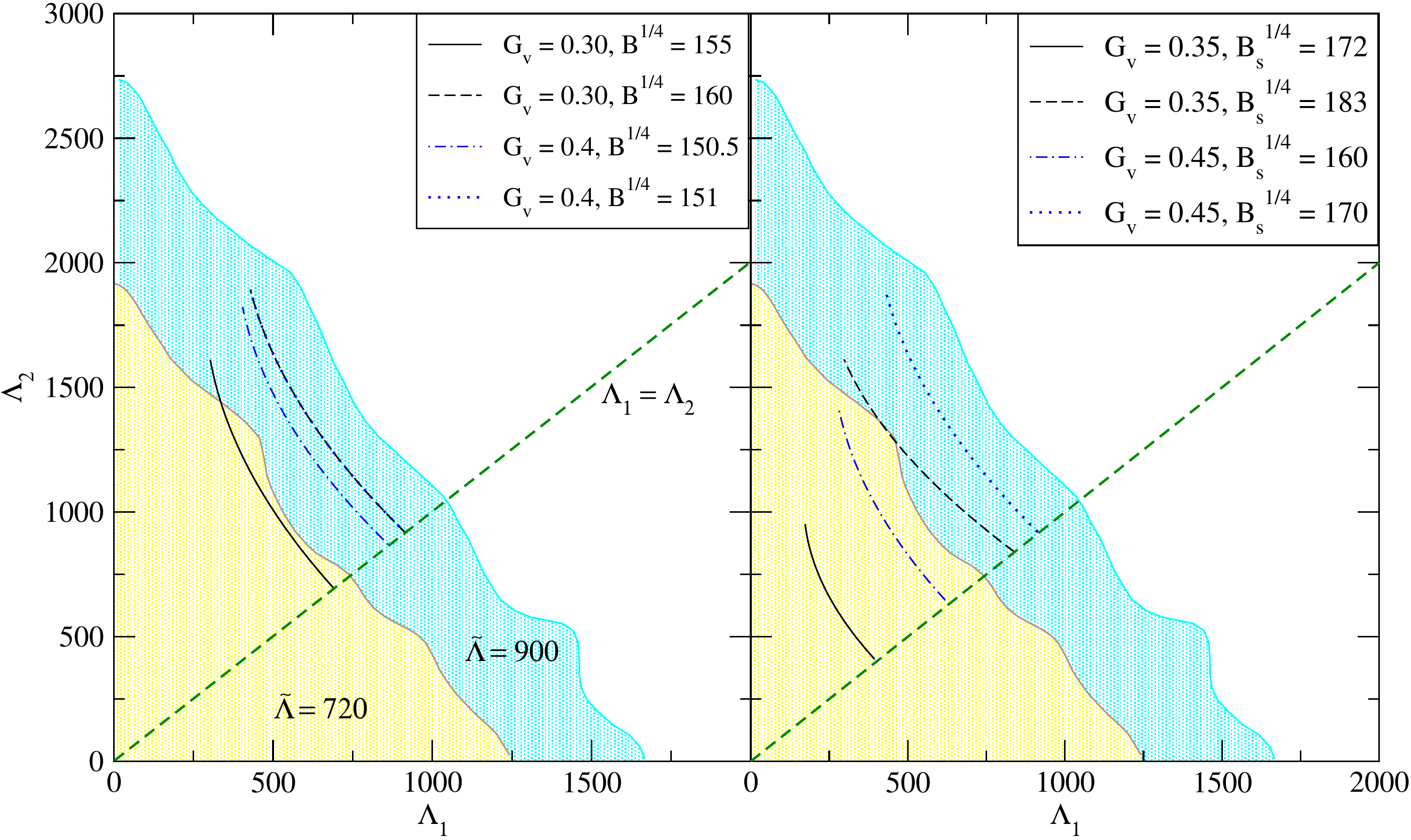}
\caption{Tidal deformabilities associated with the both components of the binary of GW170817. The shaded regions indicate the allowed region in the plane %are constraints 
from two estimates of $\tilde{\Lambda}$ from GW170817 observations.}%effective tidal deformability parameter. 
\label{fig-006}
\end{center}
\end{figure} 

\begin{table*}[t!]
\begin{center}
\caption{Properties of stars corresponding to different set of parameters. DI denotes density independent and DD represents density dependent $B$.}
\begin{tabular}{ccccccccccccc}
\hline \hline
       & $G_V$ & $B^{1/4}$ & $\mu_t$& $P_t$ & $M_{\text{max}}$ & R & $\epsilon_c$ & $R_{1.4}$ & $R_{2.08}$& $\tilde{\Lambda}$ & $\Lambda_{1.4}$ \\
     &(fm$^2$)& (MeV)  & (MeV)& (MeV/$\text{fm}^3$) & ($M_{\odot})$  & (km)  & (MeV/$\text{fm}^3$) & (km) & (km) & $q=0.73$ &  &\\         
\hline  & 0.3 & 155 & 1060 & 21 & 2.22 & 11.64 & 1136.5 & 12.65 & 12.39 & 725.1 & 596 &  \\
DI & 0.3 & 160 & 1262 & 91.4 & 2.23 & 12.06 & 1048 & 13.05 & 13.11 & 895.4 & 803 &  \\
  & 0.4 & 150.5 & 1057 & 20.5 & 2.42 & 12.32 & 993  & 12.96 & 13.22 & 861.6 & 755 &  \\
  & 0.4 & 151 & 1309 & 111 & 2.42 & 12.37 & 978  & 13.05 & 13.28 & 815.4 & 803 &  \\
\\
   & 0.35 & 172 & 1049 & 18.42 & 2.27 & 11.11 &  1264& 11.81 & 11.72 & 425.5  &342 & \\
DD & 0.35 & 183 & 1110 & 35.8 & 2.23 & 11.05  & 1283 & 12.74 & 11.72 & 756.1 &630 & \\
   & 0.45 & 160 & 1048 & 18& 2.46 & 11.89 & 1061 &  12.49 & 12.65 & 642.4 & 550 &  \\
   & 0.45 & 170 & 1198 & 66 &2.42 & 11.90 & 1186  & 13.05 & 12.98 & 895.4 & 803 &  \\
\hline
\end{tabular}
\label{tab:1}
\end{center} 
\end{table*}  

Next we examine the dependence of %represent 
tidal deformability parameter %($\Lambda$) 
on the matter model parameters. Fig. \ref{fig-006} depicts tidal deformability of both stars in binary merger scenario.  %The effective tidal deformability 
$\tilde{\Lambda}$ upper bounds are also incorporated to find our results compatibility with this parameter also. The observation of GW the maximum limit of $\tilde{\Lambda}$ is estimated as $900$ and $720$ respectively on two estimates. So it is seen that for density independent $B$, with the parameters compatible with M-R observations ($G_V = 0.3$ fm$^2$ and $B^{1/4} = 155$ MeV) reproduce $\tilde{\Lambda}$ well below the observational constraints from both the estimates. For density dependent $B$ case, all the parameter sets compatible with the M-R relation also satisfy the upper limit of $\tilde{\Lambda}$ obtained from GW observations. In right panel of this figure with density dependent $B$ mostly curves satisfy upper limit $\tilde{\Lambda}\leq720$ except with the \textcolor{black}{set of parameters $G_V=0.45$ $\text{fm}^2$, $B_s^{1/4}=170$ MeV and $G_V=0.35$ $\text{fm}^2$, $B_s^{1/4}=183$ MeV at some values.} %Next 
 %Assuming $B^{1/4}=155$ MeV as density independent and repulsive vector interaction parameter $G_V=0.3$ $\text{fm}^2$, latest limit $\tilde{\Lambda}\leq720$ is satisfied. Remaining sets of parameters are lying between this upper bound and another $\tilde{\Lambda}\leq900$.

\begin{figure} [h!]
\begin{center}
\includegraphics[width=8.7cm, keepaspectratio]{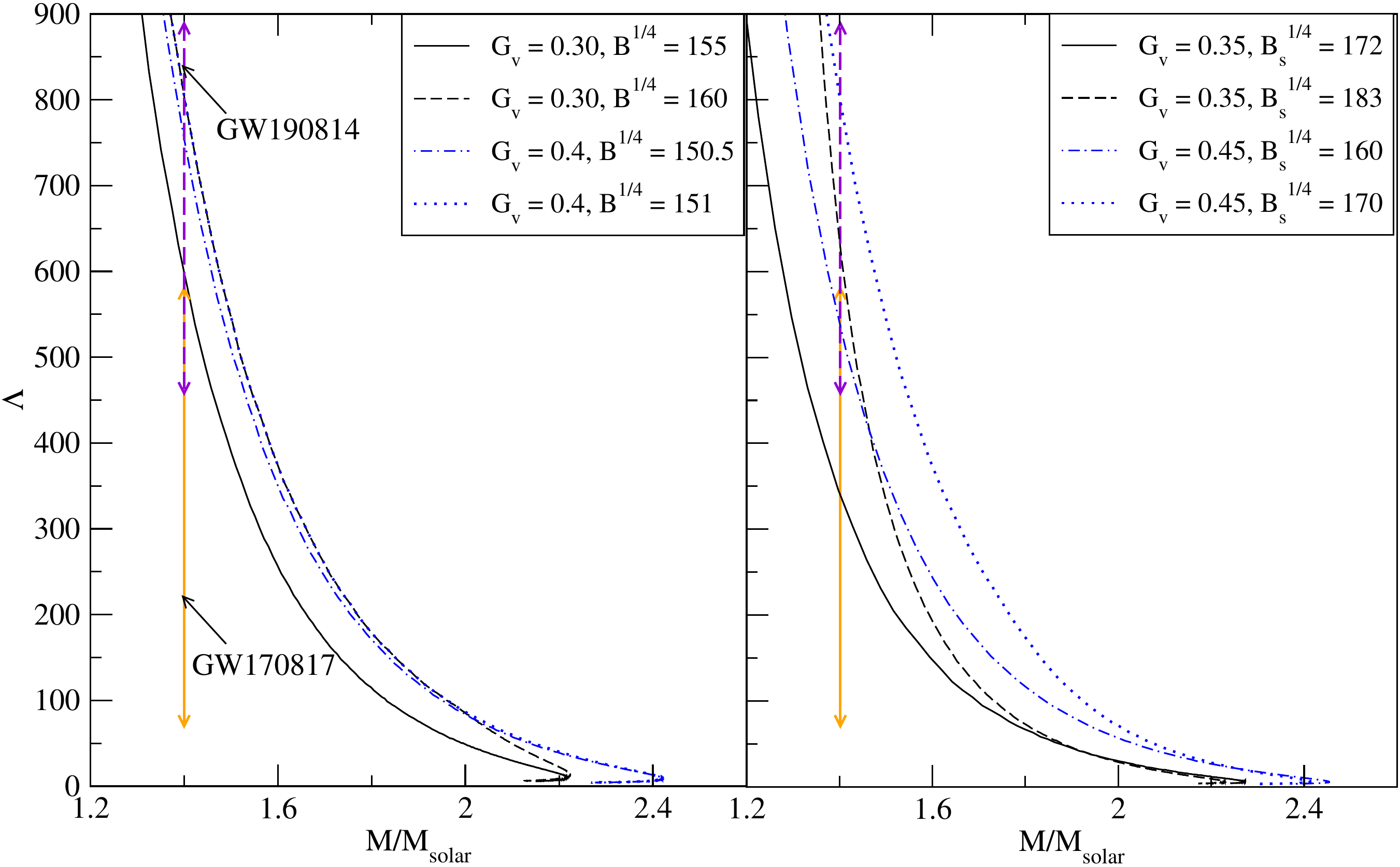}
\caption{Variation of $\Lambda$ with stellar mass. %Tidal deformability parameter $\Lambda$ for different star masses along with 
The vertical arrows show the constraints from GW190814 and GW170817 event data. %Shaded regions represents mass constraints of PSR J0952-607 and GW190814's scondary component. 
}
\label{fig-005}
\end{center}
\end{figure}

However, for density independent $B$ case the matter is too stiff to satisfy the estimated range of $\Lambda$ for $1.4~M_\odot$ star from GW170817 observation. We plot the variation of $\Lambda$ with the star mass in Fig. \ref{fig-005}. $\Lambda_{1.4}$ upper bound from GW170817 event is not satisfied with any set of parameters without density dependent $B$. For density dependent $B$ case the sets ($G_V = 0.45$ fm$^2$,$B_s^{1/4} = 160$) and ($G_V = 0.35$ fm$^2$,$B_s^{1/4} = 172$ MeV) come under the range of this estimate. This is because, with density dependent $B$ the quark phase appears earlier compared to density independent $B$ case making the matter softer even for $1.4~M_\odot$ star. Other softer sets which come within this range fail to satisfy the observed constrained to M-R relations because of higher values of $B$. For the set of parameter ($G_V=0.45$ $\text{fm}^2$,$B_s^{1/4}=170$ MeV)  $\Lambda_{1.4}$ is very high due to PT occurs at high density as shown in Fig.-(\ref{fig-002}) and star does not contain core of quark matter.

%In fig.-\ref{fig-005} tidal deformability parameter ($\Lambda$) is plotted with respect to different masses with the parameters we already considered. We have constraints on $\Lambda$ of canonical star of mass $1.4$ solar mass from GW observations.  GW190814 event's $\Lambda_{1.4}$ constraint is satisfied but maximum mass with every set of parameter is less than its lower mass bound. On the other hand $\Lambda_{1.4}$ falls within the range of GW170817 event's $\Lambda_{1.4}$ upper bound after considering $B$ as density dependent parameter and its smaller values at the surface. Larger repulsive vector interactions with lower $B$ at surface results very massive stars and also satisfying $\Lambda_{1.4}$ limits. But the lower limit of mass of secondary component of GW190814 is far from theoretical maximum mass we obtained. The set of parameter ($G_V=0.45$ $\text{fm}^2$,$B_s^{1/4}=160$ MeV) satisfies every M-R constraint as well as tidal deformability bounds.

For different parameter sets with density independent and dependent parameter $B$ we tabulated the values at the point of PT and the corresponding star properties in Table \ref{tab:1}. We have listed the parameter sets which are satisfying the M-R constraints from different observations. However, though the values of $\tilde{\Lambda}$ for all chosen parametrizations are coming under the older estimate only certain parameters sets in density dependent bag parameter scheme satisfy the recent estimate of upper limit which also reproduce $\Lambda_{1.4}$ within the estimated range from GW170817 observations. 

\begin{figure} [t!]
\begin{center}
\includegraphics[width=8.7cm, keepaspectratio]{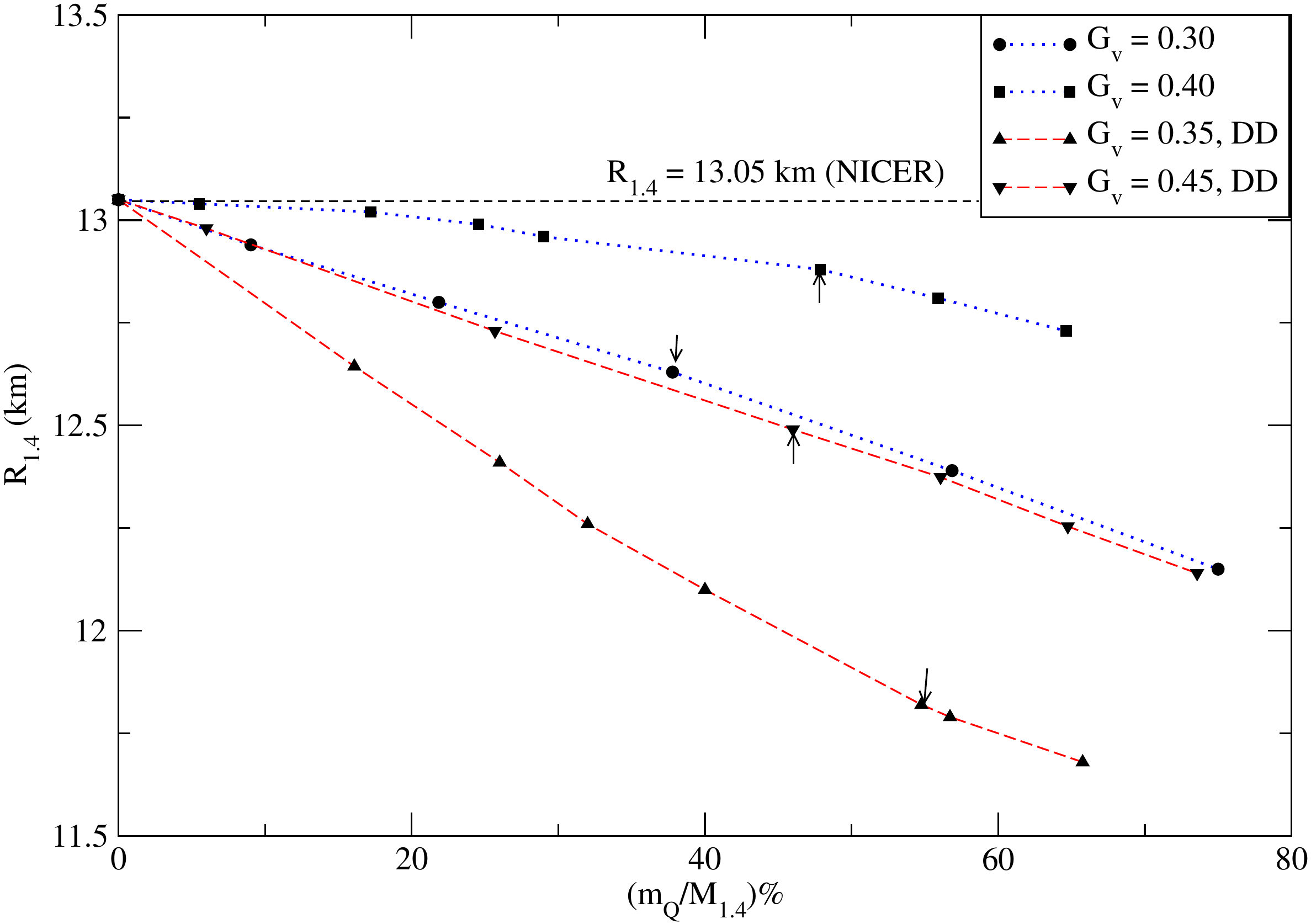}
\caption{Variation of radius of $1.4~M_\odot$ star with content of quark matter inside HS %these with 
for different parametrizations.}%repulsive interactions.
\label{fig-008}
\end{center}
\end{figure}

 As already noticed, the early appearance of SQM reduces the radius of intermediate mass star we study the behaviour of $R_{1.4}$ with the quark content of the star. We parametrize the quark content by the ratio of total SQM mass inside the star to the stellar mass. We plot $R_{1.4}$ with quark content for different parametrizations in both density independent and dependent $B$ parameter schemes in Fig. \ref{fig-008}. We see tight correlation between them for a specific value of $G_V$. Considering the minimum baryonic density for PT to occur to be $\sim0.24$ fm$^{-3}$, corresponding quark matter content is shown by arrow in the figure which shows the upper limit of quark content inside a HS. Quark matter with density dependent $B$ decreases radius more effectively than without density dependent $B$. More repulsive vector interaction causes \textcolor{black}{stiffer} EOS and results larger radius of $1.4M_\odot$.

\begin{figure} [h!]
\begin{center}
\includegraphics[width=8.7cm, keepaspectratio]{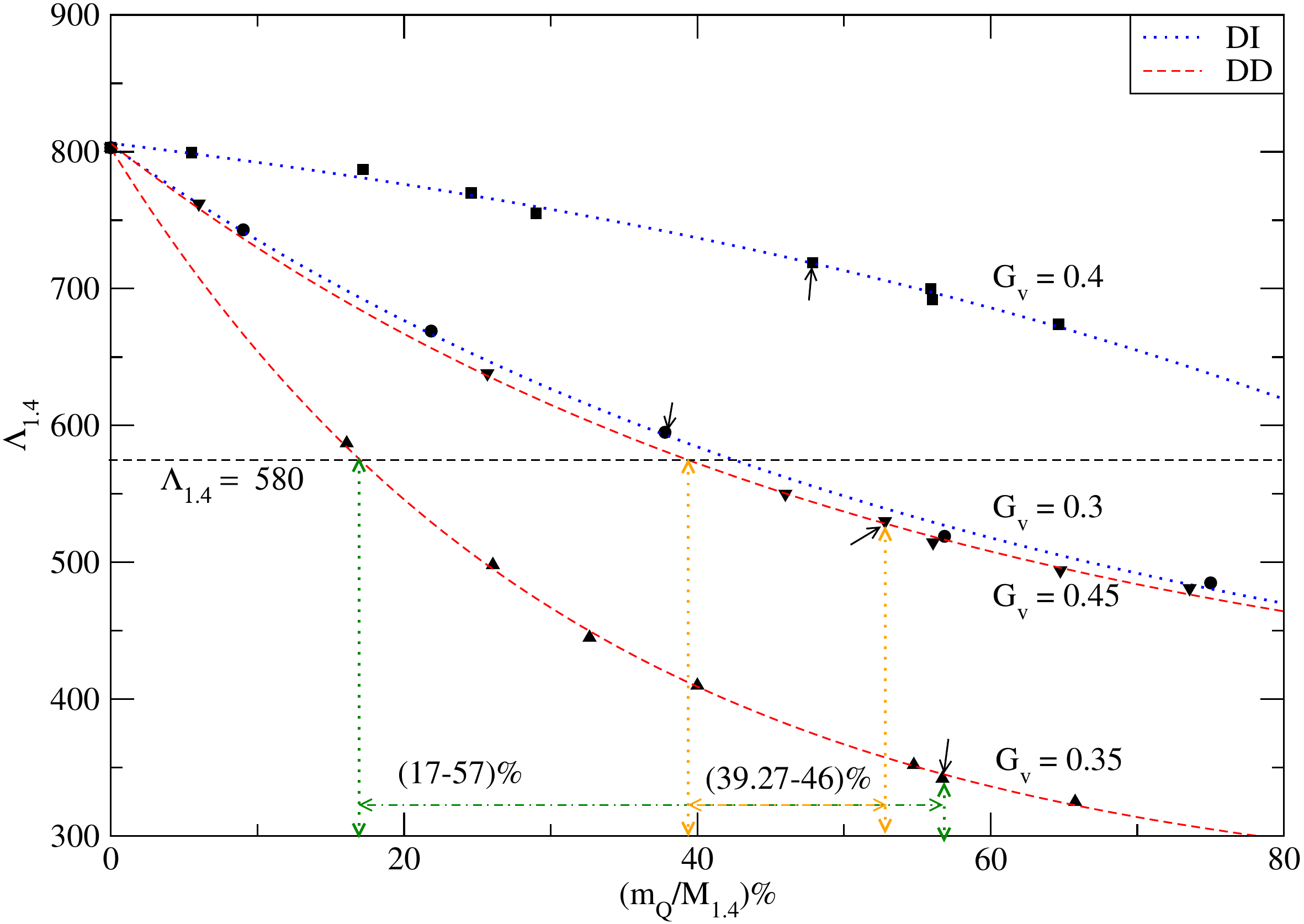}
\caption{Variation of $\Lambda$ of $1.4~M_\odot$ star with content of quark matter inside HS for different parametrizations}
\label{fig-007}
\end{center}
\end{figure}

\begin{table} [h!]
\begin{center}
\caption{Curve fitting parameters for the relation of eqn.-(\ref{eqn:e}). (std. dev.) represents maximum standard deviation. }
\begin{tabular}{cccccccccccc}
\hline \hline
      type & $G_V$   & a & b & c & $\chi^2$ & std. dev. & $R2$ \\
       of $B$& (fm$^2$) &   &   &   &          & $\%$
        & \\         
\hline
    & 0.3 & 458.12 & 1.02 & 346.67 & 0.1802 & 1.53 & 0.9987 \\
 DI & 0.4 & -98.37 & 0.987 & 904.61 & 0.1503 & 0.77 & 0.9942	 \\
 \\
    & 0.35 & 549.43 & 1.032 & 253.14 & 0.1452 & 1.07 & 0.9997 \\
 DD & 0.45 & 435.11 & 1.019 & 370.91 & 0.118 & 0.943 & 0.9993  \\
\hline
\end{tabular}
\label{tab:2}
\end{center} 
\end{table}  
\begin{table} [h!]
\begin{center}
\caption{\textcolor{black}{Curve fitting parameters for the relation of eqn.-(\ref{eqn:eq}). (std. dev.) represents maximum standard deviation.}}
\begin{tabular}{cccccccccccc}
\hline \hline
      type & $G_V$   & a & b & c & $\chi^2$ & std. dev. & $R2$ \\
       of $B$& (fm$^2$) &   &   &   &          & $\%$
        & \\         
\hline
    & 0.3 & 0.0301& -7.44 & 937.25 & 0.125 & 0.7741 & 0.9989 \\
 DI & 0.4 & -0.0136 & -1.59 & 931.24 & 0.038 & 0.434 & 0.9985	 \\
 \\
    & 0.35 & 0.0995 & -1.56 & 962 & 1.5 & 4.13 & 0.9979 \\
 DD & 0.45 & 0.044 & -8.56 & 939.6 & 0.068 & 0.575 & 0.999  \\
\hline
\end{tabular}
\label{tab:4}
\end{center} 
\end{table} 
Moreover, as the appearance of SQM softens the matter, it 
 %As we have }already noticed in previous results that appearance of quark matter 
 reduces the tidal deformability parameter. Hence $\Lambda_{1.4}$ also depends of the quark content of the star. In Fig. \ref{fig-007} we represent plot $\Lambda_{1.4}$ %effects on tidal deformability 
 with respect to content of quark matter with different parametrization with density independent and dependent $B$ parameters in hybrid star. Here $\Lambda_{1.4}$ shows strong correlation with quark content for specific values of $G_V$. We fit them by 
 \begin{equation}
\Lambda_{1.4} = ab^{-q}+c
\label{eqn:e}
\end{equation}     
where $q=\frac{m_Q}{M_{1.4}}\%$. The values of a,b and c are represented in Table \ref{tab:2}.
  The arrow indicates the same points as in Fig. \ref{fig-008}. Hence, from figure it is clear that for density independent $B$ case $\Lambda_{1.4}$ can not be less than the estimated value of that from GW170817 observations. However for density dependent $B$ case it is possible for several sets of the parameters. %Assuming lower limit of PT density from hadronic matter to quark matter (), we obtain the upper limit on content of quark matter corresponding to every $G_V$ value (indicated by 'arrow' symbol). We can change the PT point by changing the value of bag constant for a particular value of $G_V$. Early PT due to lower values of $B$ results large quark content. We get data points by changing $B$ values, the following relationship is obtained by best curve fitting
  \textcolor{black}{We obtain range $17\%$ to $57\%$ of quark matter content with density dependent bag constant and $G_V=0.35$ $\text{fm}^2$. If we assume higher repulsive vector interactions $G_V=0.45$ $\text{fm}^2$ this range of quark content becomes narrow as $39.27\%$ to $46\%$.}  
 %From the position of arrow it can be seen than for density dependent $B$ maximum quark content is lying below the upper bound $\Lambda_{1.4}\sim580$. But for the case of density independent $B$ the position of arrow is above this upper bound. 
 
\begin{figure} [h!]
\begin{center}
\includegraphics[width=8.7cm, keepaspectratio]{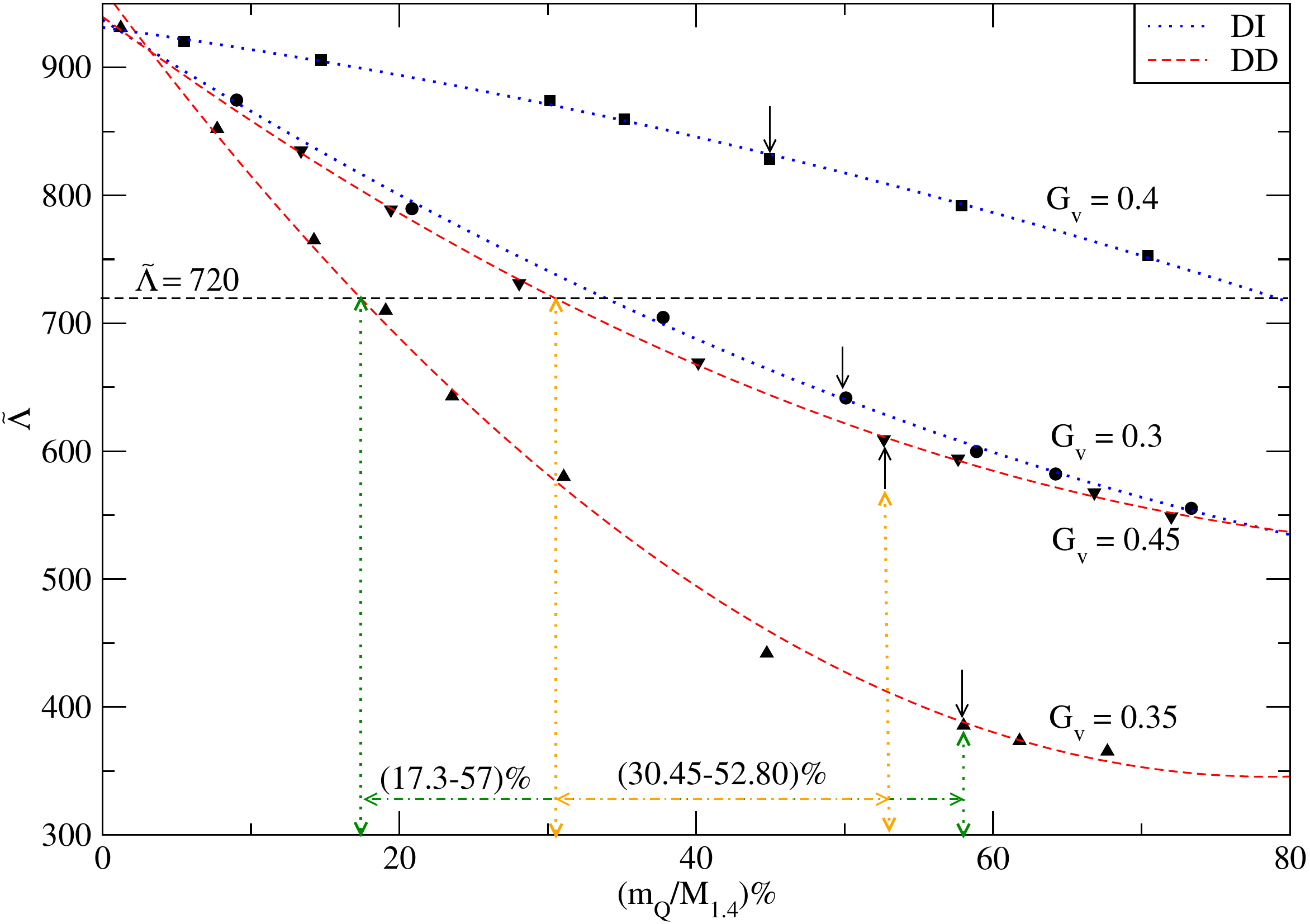}
\caption{\textcolor{black}{Variation of $\Tilde{\Lambda}$ of $1.4~M_\odot$ star with content of quark matter inside HS for different parametrizations}}
\label{fig-009}
\end{center}
\end{figure}  
\textcolor{black}{Similar to Fig. \ref{fig-007}, upper limit of effective tidal deformbility ($\Tilde{\Lambda}\leq720$) should be satisfied due to presence of quark matter. In Fig. \ref{fig-009} we represent variation of $\Tilde{\Lambda}$ with respect to content of quark matter inside the $1.4$ solar mass star. We assume primary component of mass $1.4 M_\odot$ and secondary component of mass $1.33M_\odot$. This combination of masses is lying within the range of chirp mass, mass ratio and total mass provided for GW170817 event \citep{LIGO_Virgo2017c}.  Here we find a strong correlation between $\Tilde{\Lambda}$ and content of quark matter with different repulsive vector interactions for density dependent and density independent $B$. The correlation is polynomial as
 \begin{equation}
\Tilde{\Lambda} = aq^2 + bq +c
\label{eqn:eq}
\end{equation}
where $q=\frac{m_Q}{M_{1.4}}\%$. The values of a,b and c are represented in Table \ref{tab:4}. Assuming $G_V=0.35$ $\text{fm}^2$ with density dependent bag constant we find quark content range from $17.3\%$ to $57\%$. We obtain narrow range $30.45\%$ to $52.80\%$ with higher repulsive vector interactions $G_V=0.45$ $\text{fm}^2$. Content of quark matter with density independent bag constant $G_V=0.3$ $\text{fm}^2$ also satisfies this limit. On the other hand this content of quark matter does not satisfy $\Lambda_{1.4}$ upper limit. 
Therefore from observations we get quantitative limit on the quark content inside HS which is compatible with the model of star discussed in this work. We estimate the range of quark matter content with different hadronic matter EOSs as tabulated in Table \ref{tab:5}. Parameter $G_V=0.45$ $\text{fm}^2$ does not provide single equilibrium point with GM1 EOS.}   
\begin{table} [h!]
\begin{center}
\caption{\textcolor{black}{Percentage of quark content with different hadronic matter EOSs.}}
\begin{tabular}{cccccccccc}
\hline \hline
      EOS & $G_V$   & min. & max. \\
       & (fm$^2$)  & $\%$ & $\%$ \\         
\hline
\multirow{2}{*}{DDME2} & 0.45 & 33.02 & 46.00 & \\
  & 0.35 & 12.80 & 59.72 & \\
  \hline
\multirow{2}{*}{GM1} & 0.45 & - & - & \\
  & 0.35 & 22.97 & 56.00 & \\
  \hline
\multirow{2}{*}{DD-MEX} & 0.45 & 39.27 & 46.00 & \\
 & 0.35 & 17.30 & 57.00 & \\
\hline
\end{tabular}
\label{tab:5}
\end{center} 
\end{table} 
\section{Conclusion}\label{conclusions}
We studied the HS with vector model for SQM with both density dependent as well as independent $B$. The CSs composed of only pure nucleonic matter do not fulfill their respective observational constraints. If the model of pure nucleonic matter is made to reach the lower limit of attainable maximum mass, the matter becomes too stiff to match with the upper limit of tidal deformability estimated from GW observations as well as the upper limit of \textcolor{black}{radius} near $1.4~M_\odot$ estimated from NICER observations. This motivates us to study the CS properties considering the possibility of SQM appearance inside the core of the star leading to HS. Appearance of SQM softens the matter reducing the radius. Hence to obtain the observed upper limit of radius of intermediate mass star the early appearance of SQM is favourable. This condition constraints the matter parameters. This also indicates that as early SQM appearance is favourable, the stellar properties is mainly governed by SQM model and parameters. This helps in constraining the models and parameters from astrophysical observations.

First of all, the value of $B$ is constrained for HS from the limit of minimum density of PT from hadronic matter to quark matter as PT point is highly sensitive to values of $B$. This also depends on strength of the repulsive interaction $G_V$. Lower values of $B$ and $G_V$ makes earlier PT. So this limits the minimum values for the set of $B$ and $G_V$. Also the higher values of $B$ and $G_V$ shifts the PT to higher density providing the upper limits of the set of $B$ and $G_V$ for PT to occur within the HS. From these considerations we limit $G_V$ between $0.3-0.4$ for density independent $B$ and between $0.35-0.45$ for density dependent $B$. The matter is soft if $B$ is large and $G_V$ is small. However, this is more sensible to $B$ values compared to $G_V$ values. Hence observed lower limit of maximum attainable mass and estimated $\Lambda$ from GW observations narrow down the window for $B$ and hence corresponding values of $G_V$ to obtain observed properties of CSs. From both the M-R constraints and the upper limit of $\Lambda_{1.4}$, it seems that the density independent $B$ model is not compatible with astrophysical observations.\textcolor{black}{On the hand density dependent $B$ model provide a wide range of quark matter content that satisfies both M-R constraints as well as tidal constraints obtained from observations. We estimated the common range $17.3-57\%$ and $39.27-46\%$ of quark matter which satisfy all constraints from observations with lower vector interactions and higher vector interactions respectively. Quark matter with higher vector interactions can be preferred over lower vector interaction because it satisfies M-R constraints properly. If we consider different parametrization like GM1 and DDME2 for hadronic matter EOS, we obtain almost same percentage of quark content inside the star. Quark matter having repulsive vector interactions with density dependent bag constant can be generated with NJL model as discussed in ref. \citep{2022PhRvC.105d5808C}.} After birth of CS the content of quark matter may get increase inside the core with spin-down \citep{2022MNRAS.516.1127P} which is compatible with our findings.

\begin{acknowledgments} 
V.B.T. acknowledges the funding support from a grant of the Ministry of Research, Innovation and Digitization through Project No. P4-ID-PCE-2020-0293. 
\end{acknowledgments}

%\newpage
\bibliography{references}

\begin{thebibliography}{80}%
\makeatletter
\providecommand \@ifxundefined [1]{%
 \@ifx{#1\undefined}
}%
\providecommand \@ifnum [1]{%
 \ifnum #1\expandafter \@firstoftwo
 \else \expandafter \@secondoftwo
 \fi
}%
\providecommand \@ifx [1]{%
 \ifx #1\expandafter \@firstoftwo
 \else \expandafter \@secondoftwo
 \fi
}%
\providecommand \natexlab [1]{#1}%
\providecommand \enquote  [1]{``#1''}%
\providecommand \bibnamefont  [1]{#1}%
\providecommand \bibfnamefont [1]{#1}%
\providecommand \citenamefont [1]{#1}%
\providecommand \href@noop [0]{\@secondoftwo}%
\providecommand \href [0]{\begingroup \@sanitize@url \@href}%
\providecommand \@href[1]{\@@startlink{#1}\@@href}%
\providecommand \@@href[1]{\endgroup#1\@@endlink}%
\providecommand \@sanitize@url [0]{\catcode `\\12\catcode `\$12\catcode
  `\&12\catcode `\#12\catcode `\^12\catcode `\_12\catcode `\%12\relax}%
\providecommand \@@startlink[1]{}%
\providecommand \@@endlink[0]{}%
\providecommand \url  [0]{\begingroup\@sanitize@url \@url }%
\providecommand \@url [1]{\endgroup\@href {#1}{\urlprefix }}%
\providecommand \urlprefix  [0]{URL }%
\providecommand \Eprint [0]{\href }%
\providecommand \doibase [0]{http://dx.doi.org/}%
\providecommand \selectlanguage [0]{\@gobble}%
\providecommand \bibinfo  [0]{\@secondoftwo}%
\providecommand \bibfield  [0]{\@secondoftwo}%
\providecommand \translation [1]{[#1]}%
\providecommand \BibitemOpen [0]{}%
\providecommand \bibitemStop [0]{}%
\providecommand \bibitemNoStop [0]{.\EOS\space}%
\providecommand \EOS [0]{\spacefactor3000\relax}%
\providecommand \BibitemShut  [1]{\csname bibitem#1\endcsname}%
\let\auto@bib@innerbib\@empty
%</preamble>
\bibitem [{\citenamefont {{Glendenning}}(1996)}]{1996cost.book.....G}%
  \BibitemOpen
  \bibfield  {author} {\bibinfo {author} {\bibfnamefont {N.~K.}\ \bibnamefont
  {{Glendenning}}},\ }\href@noop {} {\emph {\bibinfo {title} {{Compact
  Stars}}}}\ (\bibinfo {year} {1996})\BibitemShut {NoStop}%
\bibitem [{\citenamefont {{Antoniadis}}\ \emph {et~al.}(2013)\citenamefont
  {{Antoniadis}}, \citenamefont {{Freire}}, \citenamefont {{Wex}},
  \citenamefont {{Tauris}}, \citenamefont {{Lynch}}, \citenamefont {{van
  Kerkwijk}}, \citenamefont {{Kramer}}, \citenamefont {{Bassa}}, \citenamefont
  {{Dhillon}}, \citenamefont {{Driebe}}, \citenamefont {{Hessels}},
  \citenamefont {{Kaspi}}, \citenamefont {{Kondratiev}}, \citenamefont
  {{Langer}}, \citenamefont {{Marsh}}, \citenamefont {{McLaughlin}},
  \citenamefont {{Pennucci}}, \citenamefont {{Ransom}}, \citenamefont
  {{Stairs}}, \citenamefont {{van Leeuwen}}, \citenamefont {{Verbiest}},\ and\
  \citenamefont {{Whelan}}}]{2013Sci...340..448A}%
  \BibitemOpen
  \bibfield  {author} {\bibinfo {author} {\bibfnamefont {J.}~\bibnamefont
  {{Antoniadis}}}, \bibinfo {author} {\bibfnamefont {P.~C.~C.}\ \bibnamefont
  {{Freire}}}, \bibinfo {author} {\bibfnamefont {N.}~\bibnamefont {{Wex}}},
  \bibinfo {author} {\bibfnamefont {T.~M.}\ \bibnamefont {{Tauris}}}, \bibinfo
  {author} {\bibfnamefont {R.~S.}\ \bibnamefont {{Lynch}}}, \bibinfo {author}
  {\bibfnamefont {M.~H.}\ \bibnamefont {{van Kerkwijk}}}, \bibinfo {author}
  {\bibfnamefont {M.}~\bibnamefont {{Kramer}}}, \bibinfo {author}
  {\bibfnamefont {C.}~\bibnamefont {{Bassa}}}, \bibinfo {author} {\bibfnamefont
  {V.~S.}\ \bibnamefont {{Dhillon}}}, \bibinfo {author} {\bibfnamefont
  {T.}~\bibnamefont {{Driebe}}}, \bibinfo {author} {\bibfnamefont {J.~W.~T.}\
  \bibnamefont {{Hessels}}}, \bibinfo {author} {\bibfnamefont {V.~M.}\
  \bibnamefont {{Kaspi}}}, \bibinfo {author} {\bibfnamefont {V.~I.}\
  \bibnamefont {{Kondratiev}}}, \bibinfo {author} {\bibfnamefont
  {N.}~\bibnamefont {{Langer}}}, \bibinfo {author} {\bibfnamefont {T.~R.}\
  \bibnamefont {{Marsh}}}, \bibinfo {author} {\bibfnamefont {M.~A.}\
  \bibnamefont {{McLaughlin}}}, \bibinfo {author} {\bibfnamefont {T.~T.}\
  \bibnamefont {{Pennucci}}}, \bibinfo {author} {\bibfnamefont {S.~M.}\
  \bibnamefont {{Ransom}}}, \bibinfo {author} {\bibfnamefont {I.~H.}\
  \bibnamefont {{Stairs}}}, \bibinfo {author} {\bibfnamefont {J.}~\bibnamefont
  {{van Leeuwen}}}, \bibinfo {author} {\bibfnamefont {J.~P.~W.}\ \bibnamefont
  {{Verbiest}}}, \ and\ \bibinfo {author} {\bibfnamefont {D.~G.}\ \bibnamefont
  {{Whelan}}},\ }\href {\doibase 10.1126/science.1233232} {\bibfield  {journal}
  {\bibinfo  {journal} {\science}\ }\textbf {\bibinfo {volume} {340}},\
  \bibinfo {pages} {448} (\bibinfo {year} {2013})},\ \Eprint
  {http://arxiv.org/abs/1304.6875} {arXiv:1304.6875 [astro-ph.HE]} \BibitemShut
  {NoStop}%
\bibitem [{\citenamefont {{Cromartie}}\ \emph {et~al.}(2020)\citenamefont
  {{Cromartie}}, \citenamefont {{Fonseca}}, \citenamefont {{Ransom}},
  \citenamefont {{Demorest}}, \citenamefont {{Arzoumanian}}, \citenamefont
  {{Blumer}}, \citenamefont {{Brook}}, \citenamefont {{DeCesar}}, \citenamefont
  {{Dolch}}, \citenamefont {{Ellis}}, \citenamefont {{Ferdman}}, \citenamefont
  {{Ferrara}}, \citenamefont {{Garver-Daniels}}, \citenamefont {{Gentile}},
  \citenamefont {{Jones}}, \citenamefont {{Lam}}, \citenamefont {{Lorimer}},
  \citenamefont {{Lynch}}, \citenamefont {{McLaughlin}}, \citenamefont {{Ng}},
  \citenamefont {{Nice}}, \citenamefont {{Pennucci}}, \citenamefont
  {{Spiewak}}, \citenamefont {{Stairs}}, \citenamefont {{Stovall}},
  \citenamefont {{Swiggum}},\ and\ \citenamefont
  {{Zhu}}}]{2020NatAs...4...72C}%
  \BibitemOpen
  \bibfield  {author} {\bibinfo {author} {\bibfnamefont {H.~T.}\ \bibnamefont
  {{Cromartie}}}, \bibinfo {author} {\bibfnamefont {E.}~\bibnamefont
  {{Fonseca}}}, \bibinfo {author} {\bibfnamefont {S.~M.}\ \bibnamefont
  {{Ransom}}}, \bibinfo {author} {\bibfnamefont {P.~B.}\ \bibnamefont
  {{Demorest}}}, \bibinfo {author} {\bibfnamefont {Z.}~\bibnamefont
  {{Arzoumanian}}}, \bibinfo {author} {\bibfnamefont {H.}~\bibnamefont
  {{Blumer}}}, \bibinfo {author} {\bibfnamefont {P.~R.}\ \bibnamefont
  {{Brook}}}, \bibinfo {author} {\bibfnamefont {M.~E.}\ \bibnamefont
  {{DeCesar}}}, \bibinfo {author} {\bibfnamefont {T.}~\bibnamefont {{Dolch}}},
  \bibinfo {author} {\bibfnamefont {J.~A.}\ \bibnamefont {{Ellis}}}, \bibinfo
  {author} {\bibfnamefont {R.~D.}\ \bibnamefont {{Ferdman}}}, \bibinfo {author}
  {\bibfnamefont {E.~C.}\ \bibnamefont {{Ferrara}}}, \bibinfo {author}
  {\bibfnamefont {N.}~\bibnamefont {{Garver-Daniels}}}, \bibinfo {author}
  {\bibfnamefont {P.~A.}\ \bibnamefont {{Gentile}}}, \bibinfo {author}
  {\bibfnamefont {M.~L.}\ \bibnamefont {{Jones}}}, \bibinfo {author}
  {\bibfnamefont {M.~T.}\ \bibnamefont {{Lam}}}, \bibinfo {author}
  {\bibfnamefont {D.~R.}\ \bibnamefont {{Lorimer}}}, \bibinfo {author}
  {\bibfnamefont {R.~S.}\ \bibnamefont {{Lynch}}}, \bibinfo {author}
  {\bibfnamefont {M.~A.}\ \bibnamefont {{McLaughlin}}}, \bibinfo {author}
  {\bibfnamefont {C.}~\bibnamefont {{Ng}}}, \bibinfo {author} {\bibfnamefont
  {D.~J.}\ \bibnamefont {{Nice}}}, \bibinfo {author} {\bibfnamefont {T.~T.}\
  \bibnamefont {{Pennucci}}}, \bibinfo {author} {\bibfnamefont
  {R.}~\bibnamefont {{Spiewak}}}, \bibinfo {author} {\bibfnamefont {I.~H.}\
  \bibnamefont {{Stairs}}}, \bibinfo {author} {\bibfnamefont {K.}~\bibnamefont
  {{Stovall}}}, \bibinfo {author} {\bibfnamefont {J.~K.}\ \bibnamefont
  {{Swiggum}}}, \ and\ \bibinfo {author} {\bibfnamefont {W.~W.}\ \bibnamefont
  {{Zhu}}},\ }\href {\doibase 10.1038/s41550-019-0880-2} {\bibfield  {journal}
  {\bibinfo  {journal} {\Natureastro}\ }\textbf {\bibinfo {volume} {4}},\
  \bibinfo {pages} {72} (\bibinfo {year} {2020})},\ \Eprint
  {http://arxiv.org/abs/1904.06759} {arXiv:1904.06759 [astro-ph.HE]}
  \BibitemShut {NoStop}%
\bibitem [{\citenamefont {{Fonseca}}\ \emph {et~al.}(2021)\citenamefont
  {{Fonseca}}, \citenamefont {{Cromartie}}, \citenamefont {{Pennucci}},
  \citenamefont {{Ray}}, \citenamefont {{Kirichenko}}, \citenamefont
  {{Ransom}}, \citenamefont {{Demorest}}, \citenamefont {{Stairs}},
  \citenamefont {{Arzoumanian}}, \citenamefont {{Guillemot}}, \citenamefont
  {{Parthasarathy}}, \citenamefont {{Kerr}}, \citenamefont {{Cognard}},
  \citenamefont {{Baker}}, \citenamefont {{Blumer}}, \citenamefont {{Brook}},
  \citenamefont {{DeCesar}}, \citenamefont {{Dolch}}, \citenamefont {{Dong}},
  \citenamefont {{Ferrara}}, \citenamefont {{Fiore}}, \citenamefont
  {{Garver-Daniels}}, \citenamefont {{Good}}, \citenamefont {{Jennings}},
  \citenamefont {{Jones}}, \citenamefont {{Kaspi}}, \citenamefont {{Lam}},
  \citenamefont {{Lorimer}}, \citenamefont {{Luo}}, \citenamefont {{McEwen}},
  \citenamefont {{McKee}}, \citenamefont {{McLaughlin}}, \citenamefont
  {{McMann}}, \citenamefont {{Meyers}}, \citenamefont {{Naidu}}, \citenamefont
  {{Ng}}, \citenamefont {{Nice}}, \citenamefont {{Pol}}, \citenamefont
  {{Radovan}}, \citenamefont {{Shapiro-Albert}}, \citenamefont {{Tan}},
  \citenamefont {{Tendulkar}}, \citenamefont {{Swiggum}}, \citenamefont
  {{Wahl}},\ and\ \citenamefont {{Zhu}}}]{2021ApJ...915L..12F}%
  \BibitemOpen
  \bibfield  {author} {\bibinfo {author} {\bibfnamefont {E.}~\bibnamefont
  {{Fonseca}}}, \bibinfo {author} {\bibfnamefont {H.~T.}\ \bibnamefont
  {{Cromartie}}}, \bibinfo {author} {\bibfnamefont {T.~T.}\ \bibnamefont
  {{Pennucci}}}, \bibinfo {author} {\bibfnamefont {P.~S.}\ \bibnamefont
  {{Ray}}}, \bibinfo {author} {\bibfnamefont {A.~Y.}\ \bibnamefont
  {{Kirichenko}}}, \bibinfo {author} {\bibfnamefont {S.~M.}\ \bibnamefont
  {{Ransom}}}, \bibinfo {author} {\bibfnamefont {P.~B.}\ \bibnamefont
  {{Demorest}}}, \bibinfo {author} {\bibfnamefont {I.~H.}\ \bibnamefont
  {{Stairs}}}, \bibinfo {author} {\bibfnamefont {Z.}~\bibnamefont
  {{Arzoumanian}}}, \bibinfo {author} {\bibfnamefont {L.}~\bibnamefont
  {{Guillemot}}}, \bibinfo {author} {\bibfnamefont {A.}~\bibnamefont
  {{Parthasarathy}}}, \bibinfo {author} {\bibfnamefont {M.}~\bibnamefont
  {{Kerr}}}, \bibinfo {author} {\bibfnamefont {I.}~\bibnamefont {{Cognard}}},
  \bibinfo {author} {\bibfnamefont {P.~T.}\ \bibnamefont {{Baker}}}, \bibinfo
  {author} {\bibfnamefont {H.}~\bibnamefont {{Blumer}}}, \bibinfo {author}
  {\bibfnamefont {P.~R.}\ \bibnamefont {{Brook}}}, \bibinfo {author}
  {\bibfnamefont {M.}~\bibnamefont {{DeCesar}}}, \bibinfo {author}
  {\bibfnamefont {T.}~\bibnamefont {{Dolch}}}, \bibinfo {author} {\bibfnamefont
  {F.~A.}\ \bibnamefont {{Dong}}}, \bibinfo {author} {\bibfnamefont {E.~C.}\
  \bibnamefont {{Ferrara}}}, \bibinfo {author} {\bibfnamefont {W.}~\bibnamefont
  {{Fiore}}}, \bibinfo {author} {\bibfnamefont {N.}~\bibnamefont
  {{Garver-Daniels}}}, \bibinfo {author} {\bibfnamefont {D.~C.}\ \bibnamefont
  {{Good}}}, \bibinfo {author} {\bibfnamefont {R.}~\bibnamefont {{Jennings}}},
  \bibinfo {author} {\bibfnamefont {M.~L.}\ \bibnamefont {{Jones}}}, \bibinfo
  {author} {\bibfnamefont {V.~M.}\ \bibnamefont {{Kaspi}}}, \bibinfo {author}
  {\bibfnamefont {M.~T.}\ \bibnamefont {{Lam}}}, \bibinfo {author}
  {\bibfnamefont {D.~R.}\ \bibnamefont {{Lorimer}}}, \bibinfo {author}
  {\bibfnamefont {J.}~\bibnamefont {{Luo}}}, \bibinfo {author} {\bibfnamefont
  {A.}~\bibnamefont {{McEwen}}}, \bibinfo {author} {\bibfnamefont {J.~W.}\
  \bibnamefont {{McKee}}}, \bibinfo {author} {\bibfnamefont {M.~A.}\
  \bibnamefont {{McLaughlin}}}, \bibinfo {author} {\bibfnamefont
  {N.}~\bibnamefont {{McMann}}}, \bibinfo {author} {\bibfnamefont {B.~W.}\
  \bibnamefont {{Meyers}}}, \bibinfo {author} {\bibfnamefont {A.}~\bibnamefont
  {{Naidu}}}, \bibinfo {author} {\bibfnamefont {C.}~\bibnamefont {{Ng}}},
  \bibinfo {author} {\bibfnamefont {D.~J.}\ \bibnamefont {{Nice}}}, \bibinfo
  {author} {\bibfnamefont {N.}~\bibnamefont {{Pol}}}, \bibinfo {author}
  {\bibfnamefont {H.~A.}\ \bibnamefont {{Radovan}}}, \bibinfo {author}
  {\bibfnamefont {B.}~\bibnamefont {{Shapiro-Albert}}}, \bibinfo {author}
  {\bibfnamefont {C.~M.}\ \bibnamefont {{Tan}}}, \bibinfo {author}
  {\bibfnamefont {S.~P.}\ \bibnamefont {{Tendulkar}}}, \bibinfo {author}
  {\bibfnamefont {J.~K.}\ \bibnamefont {{Swiggum}}}, \bibinfo {author}
  {\bibfnamefont {H.~M.}\ \bibnamefont {{Wahl}}}, \ and\ \bibinfo {author}
  {\bibfnamefont {W.~W.}\ \bibnamefont {{Zhu}}},\ }\href {\doibase
  10.3847/2041-8213/ac03b8} {\bibfield  {journal} {\bibinfo  {journal} {\apjl}\
  }\textbf {\bibinfo {volume} {915}},\ \bibinfo {eid} {L12} (\bibinfo {year}
  {2021})},\ \Eprint {http://arxiv.org/abs/2104.00880} {arXiv:2104.00880
  [astro-ph.HE]} \BibitemShut {NoStop}%
\bibitem [{\citenamefont {{Miller}}\ \emph {et~al.}(2021)\citenamefont
  {{Miller}}, \citenamefont {{Lamb}}, \citenamefont {{Dittmann}}, \citenamefont
  {{Bogdanov}}, \citenamefont {{Arzoumanian}}, \citenamefont {{Gendreau}},
  \citenamefont {{Guillot}}, \citenamefont {{Ho}}, \citenamefont {{Lattimer}},
  \citenamefont {{Loewenstein}}, \citenamefont {{Morsink}}, \citenamefont
  {{Ray}}, \citenamefont {{Wolff}}, \citenamefont {{Baker}}, \citenamefont
  {{Cazeau}}, \citenamefont {{Manthripragada}}, \citenamefont {{Markwardt}},
  \citenamefont {{Okajima}}, \citenamefont {{Pollard}}, \citenamefont
  {{Cognard}}, \citenamefont {{Cromartie}}, \citenamefont {{Fonseca}},
  \citenamefont {{Guillemot}}, \citenamefont {{Kerr}}, \citenamefont
  {{Parthasarathy}}, \citenamefont {{Pennucci}}, \citenamefont {{Ransom}},\
  and\ \citenamefont {{Stairs}}}]{2021ApJ...918L..28M}%
  \BibitemOpen
  \bibfield  {author} {\bibinfo {author} {\bibfnamefont {M.~C.}\ \bibnamefont
  {{Miller}}}, \bibinfo {author} {\bibfnamefont {F.~K.}\ \bibnamefont
  {{Lamb}}}, \bibinfo {author} {\bibfnamefont {A.~J.}\ \bibnamefont
  {{Dittmann}}}, \bibinfo {author} {\bibfnamefont {S.}~\bibnamefont
  {{Bogdanov}}}, \bibinfo {author} {\bibfnamefont {Z.}~\bibnamefont
  {{Arzoumanian}}}, \bibinfo {author} {\bibfnamefont {K.~C.}\ \bibnamefont
  {{Gendreau}}}, \bibinfo {author} {\bibfnamefont {S.}~\bibnamefont
  {{Guillot}}}, \bibinfo {author} {\bibfnamefont {W.~C.~G.}\ \bibnamefont
  {{Ho}}}, \bibinfo {author} {\bibfnamefont {J.~M.}\ \bibnamefont
  {{Lattimer}}}, \bibinfo {author} {\bibfnamefont {M.}~\bibnamefont
  {{Loewenstein}}}, \bibinfo {author} {\bibfnamefont {S.~M.}\ \bibnamefont
  {{Morsink}}}, \bibinfo {author} {\bibfnamefont {P.~S.}\ \bibnamefont
  {{Ray}}}, \bibinfo {author} {\bibfnamefont {M.~T.}\ \bibnamefont {{Wolff}}},
  \bibinfo {author} {\bibfnamefont {C.~L.}\ \bibnamefont {{Baker}}}, \bibinfo
  {author} {\bibfnamefont {T.}~\bibnamefont {{Cazeau}}}, \bibinfo {author}
  {\bibfnamefont {S.}~\bibnamefont {{Manthripragada}}}, \bibinfo {author}
  {\bibfnamefont {C.~B.}\ \bibnamefont {{Markwardt}}}, \bibinfo {author}
  {\bibfnamefont {T.}~\bibnamefont {{Okajima}}}, \bibinfo {author}
  {\bibfnamefont {S.}~\bibnamefont {{Pollard}}}, \bibinfo {author}
  {\bibfnamefont {I.}~\bibnamefont {{Cognard}}}, \bibinfo {author}
  {\bibfnamefont {H.~T.}\ \bibnamefont {{Cromartie}}}, \bibinfo {author}
  {\bibfnamefont {E.}~\bibnamefont {{Fonseca}}}, \bibinfo {author}
  {\bibfnamefont {L.}~\bibnamefont {{Guillemot}}}, \bibinfo {author}
  {\bibfnamefont {M.}~\bibnamefont {{Kerr}}}, \bibinfo {author} {\bibfnamefont
  {A.}~\bibnamefont {{Parthasarathy}}}, \bibinfo {author} {\bibfnamefont
  {T.~T.}\ \bibnamefont {{Pennucci}}}, \bibinfo {author} {\bibfnamefont
  {S.}~\bibnamefont {{Ransom}}}, \ and\ \bibinfo {author} {\bibfnamefont
  {I.}~\bibnamefont {{Stairs}}},\ }\href {\doibase 10.3847/2041-8213/ac089b}
  {\bibfield  {journal} {\bibinfo  {journal} {\apjl}\ }\textbf {\bibinfo
  {volume} {918}},\ \bibinfo {eid} {L28} (\bibinfo {year} {2021})},\ \Eprint
  {http://arxiv.org/abs/2105.06979} {arXiv:2105.06979 [astro-ph.HE]}
  \BibitemShut {NoStop}%
\bibitem [{\citenamefont {{Riley}}\ \emph {et~al.}(2021)\citenamefont
  {{Riley}}, \citenamefont {{Watts}}, \citenamefont {{Ray}}, \citenamefont
  {{Bogdanov}}, \citenamefont {{Guillot}}, \citenamefont {{Morsink}},
  \citenamefont {{Bilous}}, \citenamefont {{Arzoumanian}}, \citenamefont
  {{Choudhury}}, \citenamefont {{Deneva}}, \citenamefont {{Gendreau}},
  \citenamefont {{Harding}}, \citenamefont {{Ho}}, \citenamefont {{Lattimer}},
  \citenamefont {{Loewenstein}}, \citenamefont {{Ludlam}}, \citenamefont
  {{Markwardt}}, \citenamefont {{Okajima}}, \citenamefont
  {{Prescod-Weinstein}}, \citenamefont {{Remillard}}, \citenamefont {{Wolff}},
  \citenamefont {{Fonseca}}, \citenamefont {{Cromartie}}, \citenamefont
  {{Kerr}}, \citenamefont {{Pennucci}}, \citenamefont {{Parthasarathy}},
  \citenamefont {{Ransom}}, \citenamefont {{Stairs}}, \citenamefont
  {{Guillemot}},\ and\ \citenamefont {{Cognard}}}]{2021ApJ...918L..27R}%
  \BibitemOpen
  \bibfield  {author} {\bibinfo {author} {\bibfnamefont {T.~E.}\ \bibnamefont
  {{Riley}}}, \bibinfo {author} {\bibfnamefont {A.~L.}\ \bibnamefont
  {{Watts}}}, \bibinfo {author} {\bibfnamefont {P.~S.}\ \bibnamefont {{Ray}}},
  \bibinfo {author} {\bibfnamefont {S.}~\bibnamefont {{Bogdanov}}}, \bibinfo
  {author} {\bibfnamefont {S.}~\bibnamefont {{Guillot}}}, \bibinfo {author}
  {\bibfnamefont {S.~M.}\ \bibnamefont {{Morsink}}}, \bibinfo {author}
  {\bibfnamefont {A.~V.}\ \bibnamefont {{Bilous}}}, \bibinfo {author}
  {\bibfnamefont {Z.}~\bibnamefont {{Arzoumanian}}}, \bibinfo {author}
  {\bibfnamefont {D.}~\bibnamefont {{Choudhury}}}, \bibinfo {author}
  {\bibfnamefont {J.~S.}\ \bibnamefont {{Deneva}}}, \bibinfo {author}
  {\bibfnamefont {K.~C.}\ \bibnamefont {{Gendreau}}}, \bibinfo {author}
  {\bibfnamefont {A.~K.}\ \bibnamefont {{Harding}}}, \bibinfo {author}
  {\bibfnamefont {W.~C.~G.}\ \bibnamefont {{Ho}}}, \bibinfo {author}
  {\bibfnamefont {J.~M.}\ \bibnamefont {{Lattimer}}}, \bibinfo {author}
  {\bibfnamefont {M.}~\bibnamefont {{Loewenstein}}}, \bibinfo {author}
  {\bibfnamefont {R.~M.}\ \bibnamefont {{Ludlam}}}, \bibinfo {author}
  {\bibfnamefont {C.~B.}\ \bibnamefont {{Markwardt}}}, \bibinfo {author}
  {\bibfnamefont {T.}~\bibnamefont {{Okajima}}}, \bibinfo {author}
  {\bibfnamefont {C.}~\bibnamefont {{Prescod-Weinstein}}}, \bibinfo {author}
  {\bibfnamefont {R.~A.}\ \bibnamefont {{Remillard}}}, \bibinfo {author}
  {\bibfnamefont {M.~T.}\ \bibnamefont {{Wolff}}}, \bibinfo {author}
  {\bibfnamefont {E.}~\bibnamefont {{Fonseca}}}, \bibinfo {author}
  {\bibfnamefont {H.~T.}\ \bibnamefont {{Cromartie}}}, \bibinfo {author}
  {\bibfnamefont {M.}~\bibnamefont {{Kerr}}}, \bibinfo {author} {\bibfnamefont
  {T.~T.}\ \bibnamefont {{Pennucci}}}, \bibinfo {author} {\bibfnamefont
  {A.}~\bibnamefont {{Parthasarathy}}}, \bibinfo {author} {\bibfnamefont
  {S.}~\bibnamefont {{Ransom}}}, \bibinfo {author} {\bibfnamefont
  {I.}~\bibnamefont {{Stairs}}}, \bibinfo {author} {\bibfnamefont
  {L.}~\bibnamefont {{Guillemot}}}, \ and\ \bibinfo {author} {\bibfnamefont
  {I.}~\bibnamefont {{Cognard}}},\ }\href {\doibase 10.3847/2041-8213/ac0a81}
  {\bibfield  {journal} {\bibinfo  {journal} {\apjl}\ }\textbf {\bibinfo
  {volume} {918}},\ \bibinfo {eid} {L27} (\bibinfo {year} {2021})},\ \Eprint
  {http://arxiv.org/abs/2105.06980} {arXiv:2105.06980 [astro-ph.HE]}
  \BibitemShut {NoStop}%
\bibitem [{\citenamefont {Glendenning}\ and\ \citenamefont
  {Moszkowski}(1991)}]{glendenning1991reconciliation}%
  \BibitemOpen
  \bibfield  {author} {\bibinfo {author} {\bibfnamefont {N.}~\bibnamefont
  {Glendenning}}\ and\ \bibinfo {author} {\bibfnamefont {S.}~\bibnamefont
  {Moszkowski}},\ }\href@noop {} {\bibfield  {journal} {\bibinfo  {journal}
  {Physical review letters}\ }\textbf {\bibinfo {volume} {67}},\ \bibinfo
  {pages} {2414} (\bibinfo {year} {1991})}\BibitemShut {NoStop}%
\bibitem [{\citenamefont {{Bonanno}}\ and\ \citenamefont
  {{Sedrakian}}(2012)}]{2012Astro}%
  \BibitemOpen
  \bibfield  {author} {\bibinfo {author} {\bibfnamefont {L.}~\bibnamefont
  {{Bonanno}}}\ and\ \bibinfo {author} {\bibfnamefont {A.}~\bibnamefont
  {{Sedrakian}}},\ }\href {\doibase 10.1051/0004-6361/201117832} {\bibfield
  {journal} {\bibinfo  {journal} {\aap}\ }\textbf {\bibinfo {volume} {539}},\
  \bibinfo {eid} {A16} (\bibinfo {year} {2012})},\ \Eprint
  {http://arxiv.org/abs/1108.0559} {arXiv:1108.0559 [astro-ph.SR]} \BibitemShut
  {NoStop}%
\bibitem [{\citenamefont {{Weissenborn}}\ \emph {et~al.}(2012)\citenamefont
  {{Weissenborn}}, \citenamefont {{Chatterjee}},\ and\ \citenamefont
  {{Schaffner-Bielich}}}]{2012NuPhA.881...62W}%
  \BibitemOpen
  \bibfield  {author} {\bibinfo {author} {\bibfnamefont {S.}~\bibnamefont
  {{Weissenborn}}}, \bibinfo {author} {\bibfnamefont {D.}~\bibnamefont
  {{Chatterjee}}}, \ and\ \bibinfo {author} {\bibfnamefont {J.}~\bibnamefont
  {{Schaffner-Bielich}}},\ }\href {\doibase 10.1016/j.nuclphysa.2012.02.012}
  {\bibfield  {journal} {\bibinfo  {journal} {\nphysa}\ }\textbf {\bibinfo
  {volume} {881}},\ \bibinfo {pages} {62} (\bibinfo {year} {2012})},\ \Eprint
  {http://arxiv.org/abs/1111.6049} {arXiv:1111.6049 [astro-ph.HE]} \BibitemShut
  {NoStop}%
\bibitem [{\citenamefont {{Dapo}}\ \emph {et~al.}(2010)\citenamefont {{Dapo}},
  \citenamefont {{Schaefer}},\ and\ \citenamefont
  {{Wambach}}}]{2010PhRvC..81c5803D}%
  \BibitemOpen
  \bibfield  {author} {\bibinfo {author} {\bibfnamefont {H.}~\bibnamefont
  {{Dapo}}}, \bibinfo {author} {\bibfnamefont {B.~J.}\ \bibnamefont
  {{Schaefer}}}, \ and\ \bibinfo {author} {\bibfnamefont {J.}~\bibnamefont
  {{Wambach}}},\ }\href {\doibase 10.1103/PhysRevC.81.035803} {\bibfield
  {journal} {\bibinfo  {journal} {\prc}\ }\textbf {\bibinfo {volume} {81}},\
  \bibinfo {eid} {035803} (\bibinfo {year} {2010})},\ \Eprint
  {http://arxiv.org/abs/0811.2939} {arXiv:0811.2939 [nucl-th]} \BibitemShut
  {NoStop}%
\bibitem [{\citenamefont {{Thapa}}\ \emph {et~al.}(2021)\citenamefont
  {{Thapa}}, \citenamefont {{Kumar}},\ and\ \citenamefont
  {{Sinha}}}]{2021MNRAS.507.2991T}%
  \BibitemOpen
  \bibfield  {author} {\bibinfo {author} {\bibfnamefont {V.~B.}\ \bibnamefont
  {{Thapa}}}, \bibinfo {author} {\bibfnamefont {A.}~\bibnamefont {{Kumar}}}, \
  and\ \bibinfo {author} {\bibfnamefont {M.}~\bibnamefont {{Sinha}}},\ }\href
  {\doibase 10.1093/mnras/stab2327} {\bibfield  {journal} {\bibinfo  {journal}
  {\mnras}\ }\textbf {\bibinfo {volume} {507}},\ \bibinfo {pages} {2991}
  (\bibinfo {year} {2021})},\ \Eprint {http://arxiv.org/abs/2108.04318}
  {arXiv:2108.04318 [astro-ph.HE]} \BibitemShut {NoStop}%
\bibitem [{\citenamefont {{Glendenning}}\ and\ \citenamefont
  {{Moszkowski}}(1991)}]{1991PhRvL..67.2414G}%
  \BibitemOpen
  \bibfield  {author} {\bibinfo {author} {\bibfnamefont {N.~K.}\ \bibnamefont
  {{Glendenning}}}\ and\ \bibinfo {author} {\bibfnamefont {S.~A.}\ \bibnamefont
  {{Moszkowski}}},\ }\href {\doibase 10.1103/PhysRevLett.67.2414} {\bibfield
  {journal} {\bibinfo  {journal} {\prl}\ }\textbf {\bibinfo {volume} {67}},\
  \bibinfo {pages} {2414} (\bibinfo {year} {1991})}\BibitemShut {NoStop}%
\bibitem [{\citenamefont {{Colucci}}\ and\ \citenamefont
  {{Sedrakian}}(2013)}]{2013PhRvC..87e5806C}%
  \BibitemOpen
  \bibfield  {author} {\bibinfo {author} {\bibfnamefont {G.}~\bibnamefont
  {{Colucci}}}\ and\ \bibinfo {author} {\bibfnamefont {A.}~\bibnamefont
  {{Sedrakian}}},\ }\href {\doibase 10.1103/PhysRevC.87.055806} {\bibfield
  {journal} {\bibinfo  {journal} {\prc}\ }\textbf {\bibinfo {volume} {87}},\
  \bibinfo {eid} {055806} (\bibinfo {year} {2013})},\ \Eprint
  {http://arxiv.org/abs/1302.6925} {arXiv:1302.6925 [nucl-th]} \BibitemShut
  {NoStop}%
\bibitem [{\citenamefont {{Oertel}}\ \emph {et~al.}(2016)\citenamefont
  {{Oertel}}, \citenamefont {{Gulminelli}}, \citenamefont {{Provid{\^e}ncia}},\
  and\ \citenamefont {{Raduta}}}]{2016EPJA...52...50O}%
  \BibitemOpen
  \bibfield  {author} {\bibinfo {author} {\bibfnamefont {M.}~\bibnamefont
  {{Oertel}}}, \bibinfo {author} {\bibfnamefont {F.}~\bibnamefont
  {{Gulminelli}}}, \bibinfo {author} {\bibfnamefont {C.}~\bibnamefont
  {{Provid{\^e}ncia}}}, \ and\ \bibinfo {author} {\bibfnamefont {A.~R.}\
  \bibnamefont {{Raduta}}},\ }\href {\doibase 10.1140/epja/i2016-16050-1}
  {\bibfield  {journal} {\bibinfo  {journal} {European Physical Journal A}\
  }\textbf {\bibinfo {volume} {52}},\ \bibinfo {eid} {50} (\bibinfo {year}
  {2016})},\ \Eprint {http://arxiv.org/abs/1601.00435} {arXiv:1601.00435
  [nucl-th]} \BibitemShut {NoStop}%
\bibitem [{\citenamefont {{Raduta}}\ \emph {et~al.}(2018)\citenamefont
  {{Raduta}}, \citenamefont {{Sedrakian}},\ and\ \citenamefont
  {{Weber}}}]{2018MNRAS.475.4347R}%
  \BibitemOpen
  \bibfield  {author} {\bibinfo {author} {\bibfnamefont {A.~R.}\ \bibnamefont
  {{Raduta}}}, \bibinfo {author} {\bibfnamefont {A.}~\bibnamefont
  {{Sedrakian}}}, \ and\ \bibinfo {author} {\bibfnamefont {F.}~\bibnamefont
  {{Weber}}},\ }\href {\doibase 10.1093/mnras/stx3318} {\bibfield  {journal}
  {\bibinfo  {journal} {\mnras}\ }\textbf {\bibinfo {volume} {475}},\ \bibinfo
  {pages} {4347} (\bibinfo {year} {2018})},\ \Eprint
  {http://arxiv.org/abs/1712.00584} {arXiv:1712.00584 [astro-ph.HE]}
  \BibitemShut {NoStop}%
\bibitem [{\citenamefont {{Jie Li}}\ \emph {et~al.}(2018)\citenamefont {{Jie
  Li}}, \citenamefont {{Long}},\ and\ \citenamefont
  {{Sedrakian}}}]{2018arXiv180107084J}%
  \BibitemOpen
  \bibfield  {author} {\bibinfo {author} {\bibfnamefont {J.}~\bibnamefont {{Jie
  Li}}}, \bibinfo {author} {\bibfnamefont {W.~H.}\ \bibnamefont {{Long}}}, \
  and\ \bibinfo {author} {\bibfnamefont {A.}~\bibnamefont {{Sedrakian}}},\
  }\href@noop {} {\bibfield  {journal} {\bibinfo  {journal} {arXiv e-prints}\
  ,\ \bibinfo {eid} {arXiv:1801.07084}} (\bibinfo {year} {2018})},\ \Eprint
  {http://arxiv.org/abs/1801.07084} {arXiv:1801.07084 [nucl-th]} \BibitemShut
  {NoStop}%
\bibitem [{\citenamefont {{Lopes}}\ and\ \citenamefont
  {{Menezes}}(2021)}]{2021NuPhA100922171L}%
  \BibitemOpen
  \bibfield  {author} {\bibinfo {author} {\bibfnamefont {L.~L.}\ \bibnamefont
  {{Lopes}}}\ and\ \bibinfo {author} {\bibfnamefont {D.~P.}\ \bibnamefont
  {{Menezes}}},\ }\href {\doibase 10.1016/j.nuclphysa.2021.122171} {\bibfield
  {journal} {\bibinfo  {journal} {\nphysa}\ }\textbf {\bibinfo {volume}
  {1009}},\ \bibinfo {eid} {122171} (\bibinfo {year} {2021})},\ \Eprint
  {http://arxiv.org/abs/2004.07909} {arXiv:2004.07909 [astro-ph.HE]}
  \BibitemShut {NoStop}%
\bibitem [{\citenamefont {{Lopes}}(2022)}]{2022CoTPh..74a5302L}%
  \BibitemOpen
  \bibfield  {author} {\bibinfo {author} {\bibfnamefont {L.~L.}\ \bibnamefont
  {{Lopes}}},\ }\href {\doibase 10.1088/1572-9494/ac2297} {\bibfield  {journal}
  {\bibinfo  {journal} {Communications in Theoretical Physics}\ }\textbf
  {\bibinfo {volume} {74}},\ \bibinfo {eid} {015302} (\bibinfo {year}
  {2022})},\ \Eprint {http://arxiv.org/abs/2107.02245} {arXiv:2107.02245
  [hep-ph]} \BibitemShut {NoStop}%
\bibitem [{\citenamefont {{Drago}}\ \emph {et~al.}(2014)\citenamefont
  {{Drago}}, \citenamefont {{Lavagno}}, \citenamefont {{Pagliara}},\ and\
  \citenamefont {{Pigato}}}]{2014PhRvC..90f5809D}%
  \BibitemOpen
  \bibfield  {author} {\bibinfo {author} {\bibfnamefont {A.}~\bibnamefont
  {{Drago}}}, \bibinfo {author} {\bibfnamefont {A.}~\bibnamefont {{Lavagno}}},
  \bibinfo {author} {\bibfnamefont {G.}~\bibnamefont {{Pagliara}}}, \ and\
  \bibinfo {author} {\bibfnamefont {D.}~\bibnamefont {{Pigato}}},\ }\href
  {\doibase 10.1103/PhysRevC.90.065809} {\bibfield  {journal} {\bibinfo
  {journal} {\prc}\ }\textbf {\bibinfo {volume} {90}},\ \bibinfo {eid} {065809}
  (\bibinfo {year} {2014})}\BibitemShut {NoStop}%
\bibitem [{\citenamefont {{Cai}}\ \emph {et~al.}(2015)\citenamefont {{Cai}},
  \citenamefont {{Fattoyev}}, \citenamefont {{Li}},\ and\ \citenamefont
  {{Newton}}}]{2015PhRvC..92a5802C}%
  \BibitemOpen
  \bibfield  {author} {\bibinfo {author} {\bibfnamefont {B.-J.}\ \bibnamefont
  {{Cai}}}, \bibinfo {author} {\bibfnamefont {F.~J.}\ \bibnamefont
  {{Fattoyev}}}, \bibinfo {author} {\bibfnamefont {B.-A.}\ \bibnamefont
  {{Li}}}, \ and\ \bibinfo {author} {\bibfnamefont {W.~G.}\ \bibnamefont
  {{Newton}}},\ }\href {\doibase 10.1103/PhysRevC.92.015802} {\bibfield
  {journal} {\bibinfo  {journal} {\prc}\ }\textbf {\bibinfo {volume} {92}},\
  \bibinfo {eid} {015802} (\bibinfo {year} {2015})},\ \Eprint
  {http://arxiv.org/abs/1501.01680} {arXiv:1501.01680 [nucl-th]} \BibitemShut
  {NoStop}%
\bibitem [{\citenamefont {{Li}}\ \emph {et~al.}(2018)\citenamefont {{Li}},
  \citenamefont {{Sedrakian}},\ and\ \citenamefont
  {{Weber}}}]{2018PhLB..783..234L}%
  \BibitemOpen
  \bibfield  {author} {\bibinfo {author} {\bibfnamefont {J.~J.}\ \bibnamefont
  {{Li}}}, \bibinfo {author} {\bibfnamefont {A.}~\bibnamefont {{Sedrakian}}}, \
  and\ \bibinfo {author} {\bibfnamefont {F.}~\bibnamefont {{Weber}}},\ }\href
  {\doibase 10.1016/j.physletb.2018.06.051} {\bibfield  {journal} {\bibinfo
  {journal} {Physics Letters B}\ }\textbf {\bibinfo {volume} {783}},\ \bibinfo
  {pages} {234} (\bibinfo {year} {2018})},\ \Eprint
  {http://arxiv.org/abs/1803.03661} {arXiv:1803.03661 [nucl-th]} \BibitemShut
  {NoStop}%
\bibitem [{\citenamefont {{Li}}\ and\ \citenamefont
  {{Sedrakian}}(2019)}]{2019ApJ...874L..22L}%
  \BibitemOpen
  \bibfield  {author} {\bibinfo {author} {\bibfnamefont {J.~J.}\ \bibnamefont
  {{Li}}}\ and\ \bibinfo {author} {\bibfnamefont {A.}~\bibnamefont
  {{Sedrakian}}},\ }\href {\doibase 10.3847/2041-8213/ab1090} {\bibfield
  {journal} {\bibinfo  {journal} {\apjl}\ }\textbf {\bibinfo {volume} {874}},\
  \bibinfo {eid} {L22} (\bibinfo {year} {2019})},\ \Eprint
  {http://arxiv.org/abs/1904.02006} {arXiv:1904.02006 [nucl-th]} \BibitemShut
  {NoStop}%
\bibitem [{\citenamefont {{Sedrakian}}\ \emph {et~al.}(2020)\citenamefont
  {{Sedrakian}}, \citenamefont {{Weber}},\ and\ \citenamefont
  {{Li}}}]{2020PhRvD.102d1301S}%
  \BibitemOpen
  \bibfield  {author} {\bibinfo {author} {\bibfnamefont {A.}~\bibnamefont
  {{Sedrakian}}}, \bibinfo {author} {\bibfnamefont {F.}~\bibnamefont
  {{Weber}}}, \ and\ \bibinfo {author} {\bibfnamefont {J.~J.}\ \bibnamefont
  {{Li}}},\ }\href {\doibase 10.1103/PhysRevD.102.041301} {\bibfield  {journal}
  {\bibinfo  {journal} {\prd}\ }\textbf {\bibinfo {volume} {102}},\ \bibinfo
  {eid} {041301} (\bibinfo {year} {2020})},\ \Eprint
  {http://arxiv.org/abs/2007.09683} {arXiv:2007.09683 [astro-ph.HE]}
  \BibitemShut {NoStop}%
\bibitem [{\citenamefont {{Baruah Thapa}}\ \emph {et~al.}(2020)\citenamefont
  {{Baruah Thapa}}, \citenamefont {{Sinha}}, \citenamefont {{Li}},\ and\
  \citenamefont {{Sedrakian}}}]{2020arXiv201000981B}%
  \BibitemOpen
  \bibfield  {author} {\bibinfo {author} {\bibfnamefont {V.}~\bibnamefont
  {{Baruah Thapa}}}, \bibinfo {author} {\bibfnamefont {M.}~\bibnamefont
  {{Sinha}}}, \bibinfo {author} {\bibfnamefont {J.-J.}\ \bibnamefont {{Li}}}, \
  and\ \bibinfo {author} {\bibfnamefont {A.}~\bibnamefont {{Sedrakian}}},\
  }\href@noop {} {\bibfield  {journal} {\bibinfo  {journal} {arXiv e-prints}\
  ,\ \bibinfo {eid} {arXiv:2010.00981}} (\bibinfo {year} {2020})},\ \Eprint
  {http://arxiv.org/abs/2010.00981} {arXiv:2010.00981 [hep-ph]} \BibitemShut
  {NoStop}%
\bibitem [{\citenamefont {{Mannarelli}}(2019)}]{2019arXiv190802042M}%
  \BibitemOpen
  \bibfield  {author} {\bibinfo {author} {\bibfnamefont {M.}~\bibnamefont
  {{Mannarelli}}},\ }\href@noop {} {\bibfield  {journal} {\bibinfo  {journal}
  {arXiv e-prints}\ ,\ \bibinfo {eid} {arXiv:1908.02042}} (\bibinfo {year}
  {2019})},\ \Eprint {http://arxiv.org/abs/1908.02042} {arXiv:1908.02042
  [hep-ph]} \BibitemShut {NoStop}%
\bibitem [{\citenamefont {{Thapa}}\ and\ \citenamefont
  {{Sinha}}(2020)}]{2020PhRvD.102l3007T}%
  \BibitemOpen
  \bibfield  {author} {\bibinfo {author} {\bibfnamefont {V.~B.}\ \bibnamefont
  {{Thapa}}}\ and\ \bibinfo {author} {\bibfnamefont {M.}~\bibnamefont
  {{Sinha}}},\ }\href {\doibase 10.1103/PhysRevD.102.123007} {\bibfield
  {journal} {\bibinfo  {journal} {\prd}\ }\textbf {\bibinfo {volume} {102}},\
  \bibinfo {eid} {123007} (\bibinfo {year} {2020})},\ \Eprint
  {http://arxiv.org/abs/2011.06440} {arXiv:2011.06440 [astro-ph.HE]}
  \BibitemShut {NoStop}%
\bibitem [{\citenamefont {{Haensel}}\ and\ \citenamefont
  {{Proszynski}}(1982)}]{1982ApJ...258..306H}%
  \BibitemOpen
  \bibfield  {author} {\bibinfo {author} {\bibfnamefont {P.}~\bibnamefont
  {{Haensel}}}\ and\ \bibinfo {author} {\bibfnamefont {M.}~\bibnamefont
  {{Proszynski}}},\ }\href {\doibase 10.1086/160080} {\bibfield  {journal}
  {\bibinfo  {journal} {\apj}\ }\textbf {\bibinfo {volume} {258}},\ \bibinfo
  {pages} {306} (\bibinfo {year} {1982})}\BibitemShut {NoStop}%
\bibitem [{\citenamefont {Thapa}\ \emph {et~al.}(2021)\citenamefont {Thapa},
  \citenamefont {Sinha}, \citenamefont {Li},\ and\ \citenamefont
  {Sedrakian}}]{PhysRevD.103.063004}%
  \BibitemOpen
  \bibfield  {author} {\bibinfo {author} {\bibfnamefont {V.~B.}\ \bibnamefont
  {Thapa}}, \bibinfo {author} {\bibfnamefont {M.}~\bibnamefont {Sinha}},
  \bibinfo {author} {\bibfnamefont {J.~J.}\ \bibnamefont {Li}}, \ and\ \bibinfo
  {author} {\bibfnamefont {A.}~\bibnamefont {Sedrakian}},\ }\href {\doibase
  10.1103/PhysRevD.103.063004} {\bibfield  {journal} {\bibinfo  {journal}
  {Phys. Rev. D}\ }\textbf {\bibinfo {volume} {103}},\ \bibinfo {pages}
  {063004} (\bibinfo {year} {2021})}\BibitemShut {NoStop}%
\bibitem [{\citenamefont {Schaffner}\ and\ \citenamefont
  {Mishustin}(1996)}]{PhysRevC.53.1416}%
  \BibitemOpen
  \bibfield  {author} {\bibinfo {author} {\bibfnamefont {J.}~\bibnamefont
  {Schaffner}}\ and\ \bibinfo {author} {\bibfnamefont {I.~N.}\ \bibnamefont
  {Mishustin}},\ }\href {\doibase 10.1103/PhysRevC.53.1416} {\bibfield
  {journal} {\bibinfo  {journal} {Phys. Rev. C}\ }\textbf {\bibinfo {volume}
  {53}},\ \bibinfo {pages} {1416} (\bibinfo {year} {1996})}\BibitemShut
  {NoStop}%
\bibitem [{\citenamefont {Vida\~na}\ \emph {et~al.}(2000)\citenamefont
  {Vida\~na}, \citenamefont {Polls}, \citenamefont {Ramos}, \citenamefont
  {Engvik},\ and\ \citenamefont {Hjorth-Jensen}}]{PhysRevC.62.035801}%
  \BibitemOpen
  \bibfield  {author} {\bibinfo {author} {\bibfnamefont {I.}~\bibnamefont
  {Vida\~na}}, \bibinfo {author} {\bibfnamefont {A.}~\bibnamefont {Polls}},
  \bibinfo {author} {\bibfnamefont {A.}~\bibnamefont {Ramos}}, \bibinfo
  {author} {\bibfnamefont {L.}~\bibnamefont {Engvik}}, \ and\ \bibinfo {author}
  {\bibfnamefont {M.}~\bibnamefont {Hjorth-Jensen}},\ }\href {\doibase
  10.1103/PhysRevC.62.035801} {\bibfield  {journal} {\bibinfo  {journal} {Phys.
  Rev. C}\ }\textbf {\bibinfo {volume} {62}},\ \bibinfo {pages} {035801}
  (\bibinfo {year} {2000})}\BibitemShut {NoStop}%
\bibitem [{\citenamefont {{Romani}}\ \emph {et~al.}(2022)\citenamefont
  {{Romani}}, \citenamefont {{Kandel}}, \citenamefont {{Filippenko}},
  \citenamefont {{Brink}},\ and\ \citenamefont
  {{Zheng}}}]{2022ApJ...934L..18R}%
  \BibitemOpen
  \bibfield  {author} {\bibinfo {author} {\bibfnamefont {R.~W.}\ \bibnamefont
  {{Romani}}}, \bibinfo {author} {\bibfnamefont {D.}~\bibnamefont {{Kandel}}},
  \bibinfo {author} {\bibfnamefont {A.~V.}\ \bibnamefont {{Filippenko}}},
  \bibinfo {author} {\bibfnamefont {T.~G.}\ \bibnamefont {{Brink}}}, \ and\
  \bibinfo {author} {\bibfnamefont {W.}~\bibnamefont {{Zheng}}},\ }\href
  {\doibase 10.3847/2041-8213/ac8007} {\bibfield  {journal} {\bibinfo
  {journal} {\apjl}\ }\textbf {\bibinfo {volume} {934}},\ \bibinfo {eid} {L18}
  (\bibinfo {year} {2022})},\ \Eprint {http://arxiv.org/abs/2207.05124}
  {arXiv:2207.05124 [astro-ph.HE]} \BibitemShut {NoStop}%
\bibitem [{\citenamefont {{Clevinger}}\ \emph {et~al.}(2022)\citenamefont
  {{Clevinger}}, \citenamefont {{Corkish}}, \citenamefont {{Aryal}},\ and\
  \citenamefont {{Dexheimer}}}]{2022EPJA...58...96C}%
  \BibitemOpen
  \bibfield  {author} {\bibinfo {author} {\bibfnamefont {A.}~\bibnamefont
  {{Clevinger}}}, \bibinfo {author} {\bibfnamefont {J.}~\bibnamefont
  {{Corkish}}}, \bibinfo {author} {\bibfnamefont {K.}~\bibnamefont {{Aryal}}},
  \ and\ \bibinfo {author} {\bibfnamefont {V.}~\bibnamefont {{Dexheimer}}},\
  }\href {\doibase 10.1140/epja/s10050-022-00745-3} {\bibfield  {journal}
  {\bibinfo  {journal} {European Physical Journal A}\ }\textbf {\bibinfo
  {volume} {58}},\ \bibinfo {eid} {96} (\bibinfo {year} {2022})},\ \Eprint
  {http://arxiv.org/abs/2205.00559} {arXiv:2205.00559 [astro-ph.HE]}
  \BibitemShut {NoStop}%
\bibitem [{\citenamefont {Bodmer}(1971)}]{PhysRevD.4.1601}%
  \BibitemOpen
  \bibfield  {author} {\bibinfo {author} {\bibfnamefont {A.~R.}\ \bibnamefont
  {Bodmer}},\ }\href {\doibase 10.1103/PhysRevD.4.1601} {\bibfield  {journal}
  {\bibinfo  {journal} {Phys. Rev. D}\ }\textbf {\bibinfo {volume} {4}},\
  \bibinfo {pages} {1601} (\bibinfo {year} {1971})}\BibitemShut {NoStop}%
\bibitem [{\citenamefont {Witten}(1984)}]{PhysRevD.30.272}%
  \BibitemOpen
  \bibfield  {author} {\bibinfo {author} {\bibfnamefont {E.}~\bibnamefont
  {Witten}},\ }\href {\doibase 10.1103/PhysRevD.30.272} {\bibfield  {journal}
  {\bibinfo  {journal} {Phys. Rev. D}\ }\textbf {\bibinfo {volume} {30}},\
  \bibinfo {pages} {272} (\bibinfo {year} {1984})}\BibitemShut {NoStop}%
\bibitem [{\citenamefont {Chodos}\ \emph {et~al.}(1974)\citenamefont {Chodos},
  \citenamefont {Jaffe}, \citenamefont {Johnson}, \citenamefont {Thorn},\ and\
  \citenamefont {Weisskopf}}]{PhysRevD.9.3471}%
  \BibitemOpen
  \bibfield  {author} {\bibinfo {author} {\bibfnamefont {A.}~\bibnamefont
  {Chodos}}, \bibinfo {author} {\bibfnamefont {R.~L.}\ \bibnamefont {Jaffe}},
  \bibinfo {author} {\bibfnamefont {K.}~\bibnamefont {Johnson}}, \bibinfo
  {author} {\bibfnamefont {C.~B.}\ \bibnamefont {Thorn}}, \ and\ \bibinfo
  {author} {\bibfnamefont {V.~F.}\ \bibnamefont {Weisskopf}},\ }\href {\doibase
  10.1103/PhysRevD.9.3471} {\bibfield  {journal} {\bibinfo  {journal} {Phys.
  Rev. D}\ }\textbf {\bibinfo {volume} {9}},\ \bibinfo {pages} {3471} (\bibinfo
  {year} {1974})}\BibitemShut {NoStop}%
\bibitem [{\citenamefont {{Gomes}}\ \emph {et~al.}(2019)\citenamefont
  {{Gomes}}, \citenamefont {{Char}},\ and\ \citenamefont
  {{Schramm}}}]{2019ApJ...877..139G}%
  \BibitemOpen
  \bibfield  {author} {\bibinfo {author} {\bibfnamefont {R.~O.}\ \bibnamefont
  {{Gomes}}}, \bibinfo {author} {\bibfnamefont {P.}~\bibnamefont {{Char}}}, \
  and\ \bibinfo {author} {\bibfnamefont {S.}~\bibnamefont {{Schramm}}},\ }\href
  {\doibase 10.3847/1538-4357/ab1751} {\bibfield  {journal} {\bibinfo
  {journal} {\apj}\ }\textbf {\bibinfo {volume} {877}},\ \bibinfo {eid} {139}
  (\bibinfo {year} {2019})},\ \Eprint {http://arxiv.org/abs/1806.04763}
  {arXiv:1806.04763 [nucl-th]} \BibitemShut {NoStop}%
\bibitem [{\citenamefont {{Furnstahl}}\ \emph {et~al.}(1997)\citenamefont
  {{Furnstahl}}, \citenamefont {{Serot}},\ and\ \citenamefont
  {{Tang}}}]{1997NuPhA.615..441F}%
  \BibitemOpen
  \bibfield  {author} {\bibinfo {author} {\bibfnamefont {R.~J.}\ \bibnamefont
  {{Furnstahl}}}, \bibinfo {author} {\bibfnamefont {B.~D.}\ \bibnamefont
  {{Serot}}}, \ and\ \bibinfo {author} {\bibfnamefont {H.-B.}\ \bibnamefont
  {{Tang}}},\ }\href {\doibase 10.1016/S0375-9474(96)00472-1} {\bibfield
  {journal} {\bibinfo  {journal} {\nphysa}\ }\textbf {\bibinfo {volume}
  {615}},\ \bibinfo {pages} {441} (\bibinfo {year} {1997})},\ \Eprint
  {http://arxiv.org/abs/nucl-th/9608035} {arXiv:nucl-th/9608035 [nucl-th]}
  \BibitemShut {NoStop}%
\bibitem [{\citenamefont {{Franzon}}\ \emph
  {et~al.}(2016{\natexlab{a}})\citenamefont {{Franzon}}, \citenamefont
  {{Gomes}},\ and\ \citenamefont {{Schramm}}}]{franzon2016effects}%
  \BibitemOpen
  \bibfield  {author} {\bibinfo {author} {\bibfnamefont {B.}~\bibnamefont
  {{Franzon}}}, \bibinfo {author} {\bibfnamefont {R.~O.}\ \bibnamefont
  {{Gomes}}}, \ and\ \bibinfo {author} {\bibfnamefont {S.}~\bibnamefont
  {{Schramm}}},\ }\href {\doibase 10.1093/mnras/stw1967} {\bibfield  {journal}
  {\bibinfo  {journal} {\mnras}\ }\textbf {\bibinfo {volume} {463}},\ \bibinfo
  {pages} {571} (\bibinfo {year} {2016}{\natexlab{a}})},\ \Eprint
  {http://arxiv.org/abs/1608.02845} {arXiv:1608.02845 [astro-ph.HE]}
  \BibitemShut {NoStop}%
\bibitem [{\citenamefont {{Lopes}}\ \emph
  {et~al.}(2021{\natexlab{a}})\citenamefont {{Lopes}}, \citenamefont
  {{Biesdorf}},\ and\ \citenamefont {{Menezes}}}]{2021PhyS...96f5303L}%
  \BibitemOpen
  \bibfield  {author} {\bibinfo {author} {\bibfnamefont {L.~L.}\ \bibnamefont
  {{Lopes}}}, \bibinfo {author} {\bibfnamefont {C.}~\bibnamefont {{Biesdorf}}},
  \ and\ \bibinfo {author} {\bibfnamefont {D.~P.}\ \bibnamefont {{Menezes}}},\
  }\href {\doibase 10.1088/1402-4896/abef34} {\bibfield  {journal} {\bibinfo
  {journal} {\physscr}\ }\textbf {\bibinfo {volume} {96}},\ \bibinfo {eid}
  {065303} (\bibinfo {year} {2021}{\natexlab{a}})},\ \Eprint
  {http://arxiv.org/abs/2005.13136} {arXiv:2005.13136 [hep-ph]} \BibitemShut
  {NoStop}%
\bibitem [{\citenamefont {{Sen}}\ \emph {et~al.}(2021)\citenamefont {{Sen}},
  \citenamefont {{Alam}},\ and\ \citenamefont
  {{Chaudhuri}}}]{2021JPhG...48j5201S}%
  \BibitemOpen
  \bibfield  {author} {\bibinfo {author} {\bibfnamefont {D.}~\bibnamefont
  {{Sen}}}, \bibinfo {author} {\bibfnamefont {N.}~\bibnamefont {{Alam}}}, \
  and\ \bibinfo {author} {\bibfnamefont {G.}~\bibnamefont {{Chaudhuri}}},\
  }\href {\doibase 10.1088/1361-6471/ac1713} {\bibfield  {journal} {\bibinfo
  {journal} {Journal of Physics G Nuclear Physics}\ }\textbf {\bibinfo {volume}
  {48}},\ \bibinfo {eid} {105201} (\bibinfo {year} {2021})},\ \Eprint
  {http://arxiv.org/abs/2107.08971} {arXiv:2107.08971 [nucl-th]} \BibitemShut
  {NoStop}%
\bibitem [{\citenamefont {{Alford}}\ \emph {et~al.}(2005)\citenamefont
  {{Alford}}, \citenamefont {{Braby}}, \citenamefont {{Paris}},\ and\
  \citenamefont {{Reddy}}}]{2005ApJ...629..969A}%
  \BibitemOpen
  \bibfield  {author} {\bibinfo {author} {\bibfnamefont {M.}~\bibnamefont
  {{Alford}}}, \bibinfo {author} {\bibfnamefont {M.}~\bibnamefont {{Braby}}},
  \bibinfo {author} {\bibfnamefont {M.}~\bibnamefont {{Paris}}}, \ and\
  \bibinfo {author} {\bibfnamefont {S.}~\bibnamefont {{Reddy}}},\ }\href
  {\doibase 10.1086/430902} {\bibfield  {journal} {\bibinfo  {journal} {\apj}\
  }\textbf {\bibinfo {volume} {629}},\ \bibinfo {pages} {969} (\bibinfo {year}
  {2005})},\ \Eprint {http://arxiv.org/abs/nucl-th/0411016}
  {arXiv:nucl-th/0411016 [nucl-th]} \BibitemShut {NoStop}%
\bibitem [{\citenamefont {{Parisi}}\ \emph {et~al.}(2020)\citenamefont
  {{Parisi}}, \citenamefont {{V{\'a}squez Flores}}, \citenamefont {{Lenzi}},
  \citenamefont {{Chen}},\ and\ \citenamefont
  {{Lugones}}}]{2020arXiv200914274P}%
  \BibitemOpen
  \bibfield  {author} {\bibinfo {author} {\bibfnamefont {A.}~\bibnamefont
  {{Parisi}}}, \bibinfo {author} {\bibfnamefont {C.}~\bibnamefont {{V{\'a}squez
  Flores}}}, \bibinfo {author} {\bibfnamefont {C.~H.}\ \bibnamefont {{Lenzi}}},
  \bibinfo {author} {\bibfnamefont {C.-S.}\ \bibnamefont {{Chen}}}, \ and\
  \bibinfo {author} {\bibfnamefont {G.}~\bibnamefont {{Lugones}}},\ }\href@noop
  {} {\bibfield  {journal} {\bibinfo  {journal} {arXiv e-prints}\ ,\ \bibinfo
  {eid} {arXiv:2009.14274}} (\bibinfo {year} {2020})},\ \Eprint
  {http://arxiv.org/abs/2009.14274} {arXiv:2009.14274 [astro-ph.HE]}
  \BibitemShut {NoStop}%
\bibitem [{\citenamefont {{Orsaria}}\ \emph {et~al.}(2014)\citenamefont
  {{Orsaria}}, \citenamefont {{Rodrigues}}, \citenamefont {{Weber}},\ and\
  \citenamefont {{Contrera}}}]{2014PhRvC..89a5806O}%
  \BibitemOpen
  \bibfield  {author} {\bibinfo {author} {\bibfnamefont {M.}~\bibnamefont
  {{Orsaria}}}, \bibinfo {author} {\bibfnamefont {H.}~\bibnamefont
  {{Rodrigues}}}, \bibinfo {author} {\bibfnamefont {F.}~\bibnamefont
  {{Weber}}}, \ and\ \bibinfo {author} {\bibfnamefont {G.~A.}\ \bibnamefont
  {{Contrera}}},\ }\href {\doibase 10.1103/PhysRevC.89.015806} {\bibfield
  {journal} {\bibinfo  {journal} {\prc}\ }\textbf {\bibinfo {volume} {89}},\
  \bibinfo {eid} {015806} (\bibinfo {year} {2014})},\ \Eprint
  {http://arxiv.org/abs/1308.1657} {arXiv:1308.1657 [nucl-th]} \BibitemShut
  {NoStop}%
\bibitem [{\citenamefont {{Alaverdyan}}(2022)}]{2022arXiv220800466A}%
  \BibitemOpen
  \bibfield  {author} {\bibinfo {author} {\bibfnamefont {G.~B.}\ \bibnamefont
  {{Alaverdyan}}},\ }\href@noop {} {\bibfield  {journal} {\bibinfo  {journal}
  {arXiv e-prints}\ ,\ \bibinfo {eid} {arXiv:2208.00466}} (\bibinfo {year}
  {2022})},\ \Eprint {http://arxiv.org/abs/2208.00466} {arXiv:2208.00466
  [nucl-th]} \BibitemShut {NoStop}%
\bibitem [{\citenamefont {Salinas}\ \emph {et~al.}(2019)\citenamefont
  {Salinas}, \citenamefont {Kl{\"a}hn},\ and\ \citenamefont
  {Jaikumar}}]{salinas2019strange}%
  \BibitemOpen
  \bibfield  {author} {\bibinfo {author} {\bibfnamefont {M.}~\bibnamefont
  {Salinas}}, \bibinfo {author} {\bibfnamefont {T.}~\bibnamefont {Kl{\"a}hn}},
  \ and\ \bibinfo {author} {\bibfnamefont {P.}~\bibnamefont {Jaikumar}},\
  }\href@noop {} {\bibfield  {journal} {\bibinfo  {journal} {Particles}\
  }\textbf {\bibinfo {volume} {2}},\ \bibinfo {pages} {447} (\bibinfo {year}
  {2019})}\BibitemShut {NoStop}%
\bibitem [{\citenamefont {{Louren{\c{c}}o}}\ \emph {et~al.}(2021)\citenamefont
  {{Louren{\c{c}}o}}, \citenamefont {{Lenzi}}, \citenamefont {{Dutra}},
  \citenamefont {{Ferrer}}, \citenamefont {{de la Incera}}, \citenamefont
  {{Paulucci}},\ and\ \citenamefont {{Horvath}}}]{2021PhRvD.103j3010L}%
  \BibitemOpen
  \bibfield  {author} {\bibinfo {author} {\bibfnamefont {O.}~\bibnamefont
  {{Louren{\c{c}}o}}}, \bibinfo {author} {\bibfnamefont {C.~H.}\ \bibnamefont
  {{Lenzi}}}, \bibinfo {author} {\bibfnamefont {M.}~\bibnamefont {{Dutra}}},
  \bibinfo {author} {\bibfnamefont {E.~J.}\ \bibnamefont {{Ferrer}}}, \bibinfo
  {author} {\bibfnamefont {V.}~\bibnamefont {{de la Incera}}}, \bibinfo
  {author} {\bibfnamefont {L.}~\bibnamefont {{Paulucci}}}, \ and\ \bibinfo
  {author} {\bibfnamefont {J.~E.}\ \bibnamefont {{Horvath}}},\ }\href {\doibase
  10.1103/PhysRevD.103.103010} {\bibfield  {journal} {\bibinfo  {journal}
  {\prd}\ }\textbf {\bibinfo {volume} {103}},\ \bibinfo {eid} {103010}
  (\bibinfo {year} {2021})},\ \Eprint {http://arxiv.org/abs/2104.07825}
  {arXiv:2104.07825 [astro-ph.HE]} \BibitemShut {NoStop}%
\bibitem [{\citenamefont {{Taninah}}\ \emph {et~al.}(2020)\citenamefont
  {{Taninah}}, \citenamefont {{Agbemava}}, \citenamefont {{Afanasjev}},\ and\
  \citenamefont {{Ring}}}]{2020PhLB..80035065T}%
  \BibitemOpen
  \bibfield  {author} {\bibinfo {author} {\bibfnamefont {A.}~\bibnamefont
  {{Taninah}}}, \bibinfo {author} {\bibfnamefont {S.~E.}\ \bibnamefont
  {{Agbemava}}}, \bibinfo {author} {\bibfnamefont {A.~V.}\ \bibnamefont
  {{Afanasjev}}}, \ and\ \bibinfo {author} {\bibfnamefont {P.}~\bibnamefont
  {{Ring}}},\ }\href {\doibase 10.1016/j.physletb.2019.135065} {\bibfield
  {journal} {\bibinfo  {journal} {Physics Letters B}\ }\textbf {\bibinfo
  {volume} {800}},\ \bibinfo {eid} {135065} (\bibinfo {year} {2020})},\ \Eprint
  {http://arxiv.org/abs/1910.13007} {arXiv:1910.13007 [nucl-th]} \BibitemShut
  {NoStop}%
\bibitem [{\citenamefont {Thapa}\ and\ \citenamefont
  {Sinha}(2022)}]{PhysRevC.105.015802}%
  \BibitemOpen
  \bibfield  {author} {\bibinfo {author} {\bibfnamefont {V.~B.}\ \bibnamefont
  {Thapa}}\ and\ \bibinfo {author} {\bibfnamefont {M.}~\bibnamefont {Sinha}},\
  }\href {\doibase 10.1103/PhysRevC.105.015802} {\bibfield  {journal} {\bibinfo
   {journal} {Phys. Rev. C}\ }\textbf {\bibinfo {volume} {105}},\ \bibinfo
  {pages} {015802} (\bibinfo {year} {2022})}\BibitemShut {NoStop}%
\bibitem [{\citenamefont {{Franzon}}\ \emph
  {et~al.}(2016{\natexlab{b}})\citenamefont {{Franzon}}, \citenamefont
  {{Gomes}},\ and\ \citenamefont {{Schramm}}}]{2016MNRAS.463..571F}%
  \BibitemOpen
  \bibfield  {author} {\bibinfo {author} {\bibfnamefont {B.}~\bibnamefont
  {{Franzon}}}, \bibinfo {author} {\bibfnamefont {R.~O.}\ \bibnamefont
  {{Gomes}}}, \ and\ \bibinfo {author} {\bibfnamefont {S.}~\bibnamefont
  {{Schramm}}},\ }\href {\doibase 10.1093/mnras/stw1967} {\bibfield  {journal}
  {\bibinfo  {journal} {\mnras}\ }\textbf {\bibinfo {volume} {463}},\ \bibinfo
  {pages} {571} (\bibinfo {year} {2016}{\natexlab{b}})},\ \Eprint
  {http://arxiv.org/abs/1608.02845} {arXiv:1608.02845 [astro-ph.HE]}
  \BibitemShut {NoStop}%
\bibitem [{\citenamefont {{Bhattacharyya}}\ \emph {et~al.}(2010)\citenamefont
  {{Bhattacharyya}}, \citenamefont {{Mishustin}},\ and\ \citenamefont
  {{Greiner}}}]{2010JPhG...37b5201B}%
  \BibitemOpen
  \bibfield  {author} {\bibinfo {author} {\bibfnamefont {A.}~\bibnamefont
  {{Bhattacharyya}}}, \bibinfo {author} {\bibfnamefont {I.~N.}\ \bibnamefont
  {{Mishustin}}}, \ and\ \bibinfo {author} {\bibfnamefont {W.}~\bibnamefont
  {{Greiner}}},\ }\href {\doibase 10.1088/0954-3899/37/2/025201} {\bibfield
  {journal} {\bibinfo  {journal} {Journal of Physics G Nuclear Physics}\
  }\textbf {\bibinfo {volume} {37}},\ \bibinfo {eid} {025201} (\bibinfo {year}
  {2010})},\ \Eprint {http://arxiv.org/abs/0905.0352} {arXiv:0905.0352
  [nucl-th]} \BibitemShut {NoStop}%
\bibitem [{\citenamefont {{Wu}}\ and\ \citenamefont
  {{Shen}}(2017)}]{2017PhRvC..96b5802W}%
  \BibitemOpen
  \bibfield  {author} {\bibinfo {author} {\bibfnamefont {X.~H.}\ \bibnamefont
  {{Wu}}}\ and\ \bibinfo {author} {\bibfnamefont {H.}~\bibnamefont {{Shen}}},\
  }\href {\doibase 10.1103/PhysRevC.96.025802} {\bibfield  {journal} {\bibinfo
  {journal} {\prc}\ }\textbf {\bibinfo {volume} {96}},\ \bibinfo {eid} {025802}
  (\bibinfo {year} {2017})},\ \Eprint {http://arxiv.org/abs/1708.01878}
  {arXiv:1708.01878 [nucl-th]} \BibitemShut {NoStop}%
\bibitem [{\citenamefont {{Voskresensky}}\ \emph {et~al.}(2003)\citenamefont
  {{Voskresensky}}, \citenamefont {{Yasuhira}},\ and\ \citenamefont
  {{Tatsumi}}}]{2003NuPhA.723..291V}%
  \BibitemOpen
  \bibfield  {author} {\bibinfo {author} {\bibfnamefont {D.~N.}\ \bibnamefont
  {{Voskresensky}}}, \bibinfo {author} {\bibfnamefont {M.}~\bibnamefont
  {{Yasuhira}}}, \ and\ \bibinfo {author} {\bibfnamefont {T.}~\bibnamefont
  {{Tatsumi}}},\ }\href {\doibase 10.1016/S0375-9474(03)01313-7} {\bibfield
  {journal} {\bibinfo  {journal} {\nphysa}\ }\textbf {\bibinfo {volume}
  {723}},\ \bibinfo {pages} {291} (\bibinfo {year} {2003})},\ \Eprint
  {http://arxiv.org/abs/nucl-th/0208067} {arXiv:nucl-th/0208067 [nucl-th]}
  \BibitemShut {NoStop}%
\bibitem [{\citenamefont {{Maruyama}}\ \emph {et~al.}(2007)\citenamefont
  {{Maruyama}}, \citenamefont {{Chiba}}, \citenamefont {{Schulze}},\ and\
  \citenamefont {{Tatsumi}}}]{2007PhRvD..76l3015M}%
  \BibitemOpen
  \bibfield  {author} {\bibinfo {author} {\bibfnamefont {T.}~\bibnamefont
  {{Maruyama}}}, \bibinfo {author} {\bibfnamefont {S.}~\bibnamefont {{Chiba}}},
  \bibinfo {author} {\bibfnamefont {H.-J.}\ \bibnamefont {{Schulze}}}, \ and\
  \bibinfo {author} {\bibfnamefont {T.}~\bibnamefont {{Tatsumi}}},\ }\href
  {\doibase 10.1103/PhysRevD.76.123015} {\bibfield  {journal} {\bibinfo
  {journal} {\prd}\ }\textbf {\bibinfo {volume} {76}},\ \bibinfo {eid} {123015}
  (\bibinfo {year} {2007})},\ \Eprint {http://arxiv.org/abs/0708.3277}
  {arXiv:0708.3277 [nucl-th]} \BibitemShut {NoStop}%
\bibitem [{\citenamefont {{Xia}}\ \emph {et~al.}(2020)\citenamefont {{Xia}},
  \citenamefont {{Maruyama}}, \citenamefont {{Yasutake}}, \citenamefont
  {{Tatsumi}}, \citenamefont {{Shen}},\ and\ \citenamefont
  {{Togashi}}}]{2020PhRvD.102b3031X}%
  \BibitemOpen
  \bibfield  {author} {\bibinfo {author} {\bibfnamefont {C.-J.}\ \bibnamefont
  {{Xia}}}, \bibinfo {author} {\bibfnamefont {T.}~\bibnamefont {{Maruyama}}},
  \bibinfo {author} {\bibfnamefont {N.}~\bibnamefont {{Yasutake}}}, \bibinfo
  {author} {\bibfnamefont {T.}~\bibnamefont {{Tatsumi}}}, \bibinfo {author}
  {\bibfnamefont {H.}~\bibnamefont {{Shen}}}, \ and\ \bibinfo {author}
  {\bibfnamefont {H.}~\bibnamefont {{Togashi}}},\ }\href {\doibase
  10.1103/PhysRevD.102.023031} {\bibfield  {journal} {\bibinfo  {journal}
  {\prd}\ }\textbf {\bibinfo {volume} {102}},\ \bibinfo {eid} {023031}
  (\bibinfo {year} {2020})},\ \Eprint {http://arxiv.org/abs/2005.02273}
  {arXiv:2005.02273 [hep-ph]} \BibitemShut {NoStop}%
\bibitem [{\citenamefont {{Miller}}\ \emph {et~al.}(2019)\citenamefont
  {{Miller}}, \citenamefont {{Lamb}}, \citenamefont {{Dittmann}}, \citenamefont
  {{Bogdanov}}, \citenamefont {{Arzoumanian}}, \citenamefont {{Gendreau}},
  \citenamefont {{Guillot}}, \citenamefont {{Harding}}, \citenamefont {{Ho}},
  \citenamefont {{Lattimer}}, \citenamefont {{Ludlam}}, \citenamefont
  {{Mahmoodifar}}, \citenamefont {{Morsink}}, \citenamefont {{Ray}},
  \citenamefont {{Strohmayer}}, \citenamefont {{Wood}}, \citenamefont
  {{Enoto}}, \citenamefont {{Foster}}, \citenamefont {{Okajima}}, \citenamefont
  {{Prigozhin}},\ and\ \citenamefont {{Soong}}}]{2019ApJ2}%
  \BibitemOpen
  \bibfield  {author} {\bibinfo {author} {\bibfnamefont {M.~C.}\ \bibnamefont
  {{Miller}}}, \bibinfo {author} {\bibfnamefont {F.~K.}\ \bibnamefont
  {{Lamb}}}, \bibinfo {author} {\bibfnamefont {A.~J.}\ \bibnamefont
  {{Dittmann}}}, \bibinfo {author} {\bibfnamefont {S.}~\bibnamefont
  {{Bogdanov}}}, \bibinfo {author} {\bibfnamefont {Z.}~\bibnamefont
  {{Arzoumanian}}}, \bibinfo {author} {\bibfnamefont {K.~C.}\ \bibnamefont
  {{Gendreau}}}, \bibinfo {author} {\bibfnamefont {S.}~\bibnamefont
  {{Guillot}}}, \bibinfo {author} {\bibfnamefont {A.~K.}\ \bibnamefont
  {{Harding}}}, \bibinfo {author} {\bibfnamefont {W.~C.~G.}\ \bibnamefont
  {{Ho}}}, \bibinfo {author} {\bibfnamefont {J.~M.}\ \bibnamefont
  {{Lattimer}}}, \bibinfo {author} {\bibfnamefont {R.~M.}\ \bibnamefont
  {{Ludlam}}}, \bibinfo {author} {\bibfnamefont {S.}~\bibnamefont
  {{Mahmoodifar}}}, \bibinfo {author} {\bibfnamefont {S.~M.}\ \bibnamefont
  {{Morsink}}}, \bibinfo {author} {\bibfnamefont {P.~S.}\ \bibnamefont
  {{Ray}}}, \bibinfo {author} {\bibfnamefont {T.~E.}\ \bibnamefont
  {{Strohmayer}}}, \bibinfo {author} {\bibfnamefont {K.~S.}\ \bibnamefont
  {{Wood}}}, \bibinfo {author} {\bibfnamefont {T.}~\bibnamefont {{Enoto}}},
  \bibinfo {author} {\bibfnamefont {R.}~\bibnamefont {{Foster}}}, \bibinfo
  {author} {\bibfnamefont {T.}~\bibnamefont {{Okajima}}}, \bibinfo {author}
  {\bibfnamefont {G.}~\bibnamefont {{Prigozhin}}}, \ and\ \bibinfo {author}
  {\bibfnamefont {Y.}~\bibnamefont {{Soong}}},\ }\href {\doibase
  10.3847/2041-8213/ab50c5} {\bibfield  {journal} {\bibinfo  {journal} {\apjl}\
  }\textbf {\bibinfo {volume} {887}},\ \bibinfo {eid} {L24} (\bibinfo {year}
  {2019})},\ \Eprint {http://arxiv.org/abs/1912.05705} {arXiv:1912.05705
  [astro-ph.HE]} \BibitemShut {NoStop}%
\bibitem [{\citenamefont {{Riley}}\ \emph {et~al.}(2019)\citenamefont
  {{Riley}}, \citenamefont {{Watts}}, \citenamefont {{Bogdanov}}, \citenamefont
  {{Ray}}, \citenamefont {{Ludlam}}, \citenamefont {{Guillot}}, \citenamefont
  {{Arzoumanian}}, \citenamefont {{Baker}}, \citenamefont {{Bilous}},
  \citenamefont {{Chakrabarty}}, \citenamefont {{Gendreau}}, \citenamefont
  {{Harding}}, \citenamefont {{Ho}}, \citenamefont {{Lattimer}}, \citenamefont
  {{Morsink}},\ and\ \citenamefont {{Strohmayer}}}]{2019ApJ4}%
  \BibitemOpen
  \bibfield  {author} {\bibinfo {author} {\bibfnamefont {T.~E.}\ \bibnamefont
  {{Riley}}}, \bibinfo {author} {\bibfnamefont {A.~L.}\ \bibnamefont
  {{Watts}}}, \bibinfo {author} {\bibfnamefont {S.}~\bibnamefont {{Bogdanov}}},
  \bibinfo {author} {\bibfnamefont {P.~S.}\ \bibnamefont {{Ray}}}, \bibinfo
  {author} {\bibfnamefont {R.~M.}\ \bibnamefont {{Ludlam}}}, \bibinfo {author}
  {\bibfnamefont {S.}~\bibnamefont {{Guillot}}}, \bibinfo {author}
  {\bibfnamefont {Z.}~\bibnamefont {{Arzoumanian}}}, \bibinfo {author}
  {\bibfnamefont {C.~L.}\ \bibnamefont {{Baker}}}, \bibinfo {author}
  {\bibfnamefont {A.~V.}\ \bibnamefont {{Bilous}}}, \bibinfo {author}
  {\bibfnamefont {D.}~\bibnamefont {{Chakrabarty}}}, \bibinfo {author}
  {\bibfnamefont {K.~C.}\ \bibnamefont {{Gendreau}}}, \bibinfo {author}
  {\bibfnamefont {A.~K.}\ \bibnamefont {{Harding}}}, \bibinfo {author}
  {\bibfnamefont {W.~C.~G.}\ \bibnamefont {{Ho}}}, \bibinfo {author}
  {\bibfnamefont {J.~M.}\ \bibnamefont {{Lattimer}}}, \bibinfo {author}
  {\bibfnamefont {S.~M.}\ \bibnamefont {{Morsink}}}, \ and\ \bibinfo {author}
  {\bibfnamefont {T.~E.}\ \bibnamefont {{Strohmayer}}},\ }\href {\doibase
  10.3847/2041-8213/ab481c} {\bibfield  {journal} {\bibinfo  {journal} {\apjl}\
  }\textbf {\bibinfo {volume} {887}},\ \bibinfo {eid} {L21} (\bibinfo {year}
  {2019})},\ \Eprint {http://arxiv.org/abs/1912.05702} {arXiv:1912.05702
  [astro-ph.HE]} \BibitemShut {NoStop}%
\bibitem [{\citenamefont {{Raaijmakers}}\ \emph {et~al.}(2021)\citenamefont
  {{Raaijmakers}}, \citenamefont {{Greif}}, \citenamefont {{Hebeler}},
  \citenamefont {{Hinderer}}, \citenamefont {{Nissanke}}, \citenamefont
  {{Schwenk}}, \citenamefont {{Riley}}, \citenamefont {{Watts}}, \citenamefont
  {{Lattimer}},\ and\ \citenamefont {{Ho}}}]{2021ApJ...918L..29R}%
  \BibitemOpen
  \bibfield  {author} {\bibinfo {author} {\bibfnamefont {G.}~\bibnamefont
  {{Raaijmakers}}}, \bibinfo {author} {\bibfnamefont {S.~K.}\ \bibnamefont
  {{Greif}}}, \bibinfo {author} {\bibfnamefont {K.}~\bibnamefont {{Hebeler}}},
  \bibinfo {author} {\bibfnamefont {T.}~\bibnamefont {{Hinderer}}}, \bibinfo
  {author} {\bibfnamefont {S.}~\bibnamefont {{Nissanke}}}, \bibinfo {author}
  {\bibfnamefont {A.}~\bibnamefont {{Schwenk}}}, \bibinfo {author}
  {\bibfnamefont {T.~E.}\ \bibnamefont {{Riley}}}, \bibinfo {author}
  {\bibfnamefont {A.~L.}\ \bibnamefont {{Watts}}}, \bibinfo {author}
  {\bibfnamefont {J.~M.}\ \bibnamefont {{Lattimer}}}, \ and\ \bibinfo {author}
  {\bibfnamefont {W.~C.~G.}\ \bibnamefont {{Ho}}},\ }\href {\doibase
  10.3847/2041-8213/ac089a} {\bibfield  {journal} {\bibinfo  {journal} {\apjl}\
  }\textbf {\bibinfo {volume} {918}},\ \bibinfo {eid} {L29} (\bibinfo {year}
  {2021})},\ \Eprint {http://arxiv.org/abs/2105.06981} {arXiv:2105.06981
  [astro-ph.HE]} \BibitemShut {NoStop}%
\bibitem [{\citenamefont {Abbott}\ \emph {et~al.}(2017)\citenamefont {Abbott},
  \citenamefont {Abbott}, \citenamefont {Abbott}, \citenamefont {Acernese},
  \citenamefont {Ackley} \emph {et~al.}}]{LIGO_Virgo2017c}%
  \BibitemOpen
  \bibfield  {author} {\bibinfo {author} {\bibfnamefont {B.~P.}\ \bibnamefont
  {Abbott}}, \bibinfo {author} {\bibfnamefont {R.}~\bibnamefont {Abbott}},
  \bibinfo {author} {\bibfnamefont {T.~D.}\ \bibnamefont {Abbott}}, \bibinfo
  {author} {\bibfnamefont {F.}~\bibnamefont {Acernese}}, \bibinfo {author}
  {\bibfnamefont {K.}~\bibnamefont {Ackley}},  \emph {et~al.} (\bibinfo
  {collaboration} {LIGO Scientific Collaboration and Virgo Collaboration}),\
  }\href@noop {} {\bibfield  {journal} {\bibinfo  {journal} {PhRvL}\ }\textbf
  {\bibinfo {volume} {119}},\ \bibinfo {pages} {161101} (\bibinfo {year}
  {2017})}\BibitemShut {NoStop}%
\bibitem [{\citenamefont {{Abbott}}\ \emph {et~al.}(2020)\citenamefont
  {{Abbott}}, \citenamefont {{Abbott}}, \citenamefont {{Abbott}}, \citenamefont
  {{Abraham}}, \citenamefont {{Acernese}}, \citenamefont {{Ackley}},
  \citenamefont {{Adams}}, \citenamefont {{Adhikari}}, \citenamefont {{Adya}},
  \citenamefont {{Affeldt}} \emph {et~al.}}]{2020ApJ1}%
  \BibitemOpen
  \bibfield  {author} {\bibinfo {author} {\bibfnamefont {B.~P.}\ \bibnamefont
  {{Abbott}}}, \bibinfo {author} {\bibfnamefont {R.}~\bibnamefont {{Abbott}}},
  \bibinfo {author} {\bibfnamefont {T.~D.}\ \bibnamefont {{Abbott}}}, \bibinfo
  {author} {\bibfnamefont {S.}~\bibnamefont {{Abraham}}}, \bibinfo {author}
  {\bibfnamefont {F.}~\bibnamefont {{Acernese}}}, \bibinfo {author}
  {\bibfnamefont {K.}~\bibnamefont {{Ackley}}}, \bibinfo {author}
  {\bibfnamefont {C.}~\bibnamefont {{Adams}}}, \bibinfo {author} {\bibfnamefont
  {R.~X.}\ \bibnamefont {{Adhikari}}}, \bibinfo {author} {\bibfnamefont
  {V.~B.}\ \bibnamefont {{Adya}}}, \bibinfo {author} {\bibfnamefont
  {C.}~\bibnamefont {{Affeldt}}},  \emph {et~al.},\ }\href {\doibase
  10.3847/2041-8213/ab75f5} {\bibfield  {journal} {\bibinfo  {journal} {\apjl}\
  }\textbf {\bibinfo {volume} {892}},\ \bibinfo {eid} {L3} (\bibinfo {year}
  {2020})},\ \Eprint {http://arxiv.org/abs/2001.01761} {arXiv:2001.01761
  [astro-ph.HE]} \BibitemShut {NoStop}%
\bibitem [{\citenamefont {{Radice}}\ \emph {et~al.}(2018)\citenamefont
  {{Radice}}, \citenamefont {{Perego}}, \citenamefont {{Zappa}},\ and\
  \citenamefont {{Bernuzzi}}}]{2018ApJ...852L..29R}%
  \BibitemOpen
  \bibfield  {author} {\bibinfo {author} {\bibfnamefont {D.}~\bibnamefont
  {{Radice}}}, \bibinfo {author} {\bibfnamefont {A.}~\bibnamefont {{Perego}}},
  \bibinfo {author} {\bibfnamefont {F.}~\bibnamefont {{Zappa}}}, \ and\
  \bibinfo {author} {\bibfnamefont {S.}~\bibnamefont {{Bernuzzi}}},\ }\href
  {\doibase 10.3847/2041-8213/aaa402} {\bibfield  {journal} {\bibinfo
  {journal} {\apjl}\ }\textbf {\bibinfo {volume} {852}},\ \bibinfo {eid} {L29}
  (\bibinfo {year} {2018})},\ \Eprint {http://arxiv.org/abs/1711.03647}
  {arXiv:1711.03647 [astro-ph.HE]} \BibitemShut {NoStop}%
\bibitem [{\citenamefont {Abbott}\ \emph {et~al.}(2019)\citenamefont {Abbott},
  \citenamefont {Abbott}, \citenamefont {Abbott}, \citenamefont {Abraham},
  \citenamefont {Acernese}, \citenamefont {Ackley}, \citenamefont {Adams},
  \citenamefont {Adhikari}, \citenamefont {Adya}, \citenamefont {Affeldt} \emph
  {et~al.}}]{abbott2019gwtc}%
  \BibitemOpen
  \bibfield  {author} {\bibinfo {author} {\bibfnamefont {B.}~\bibnamefont
  {Abbott}}, \bibinfo {author} {\bibfnamefont {R.}~\bibnamefont {Abbott}},
  \bibinfo {author} {\bibfnamefont {T.}~\bibnamefont {Abbott}}, \bibinfo
  {author} {\bibfnamefont {S.}~\bibnamefont {Abraham}}, \bibinfo {author}
  {\bibfnamefont {F.}~\bibnamefont {Acernese}}, \bibinfo {author}
  {\bibfnamefont {K.}~\bibnamefont {Ackley}}, \bibinfo {author} {\bibfnamefont
  {C.}~\bibnamefont {Adams}}, \bibinfo {author} {\bibfnamefont
  {R.}~\bibnamefont {Adhikari}}, \bibinfo {author} {\bibfnamefont
  {V.}~\bibnamefont {Adya}}, \bibinfo {author} {\bibfnamefont {C.}~\bibnamefont
  {Affeldt}},  \emph {et~al.},\ }\href@noop {} {\bibfield  {journal} {\bibinfo
  {journal} {Physical Review X}\ }\textbf {\bibinfo {volume} {9}},\ \bibinfo
  {pages} {031040} (\bibinfo {year} {2019})}\BibitemShut {NoStop}%
\bibitem [{\citenamefont {Abbott}\ \emph {et~al.}(2018)\citenamefont {Abbott},
  \citenamefont {Abbott}, \citenamefont {Abbott}, \citenamefont {Acernese},
  \citenamefont {Ackley}, \citenamefont {Adams}, \citenamefont {Adams},
  \citenamefont {Addesso}, \citenamefont {Adhikari}, \citenamefont {Adya} \emph
  {et~al.}}]{abbott2018gw170817}%
  \BibitemOpen
  \bibfield  {author} {\bibinfo {author} {\bibfnamefont {B.~P.}\ \bibnamefont
  {Abbott}}, \bibinfo {author} {\bibfnamefont {R.}~\bibnamefont {Abbott}},
  \bibinfo {author} {\bibfnamefont {T.}~\bibnamefont {Abbott}}, \bibinfo
  {author} {\bibfnamefont {F.}~\bibnamefont {Acernese}}, \bibinfo {author}
  {\bibfnamefont {K.}~\bibnamefont {Ackley}}, \bibinfo {author} {\bibfnamefont
  {C.}~\bibnamefont {Adams}}, \bibinfo {author} {\bibfnamefont
  {T.}~\bibnamefont {Adams}}, \bibinfo {author} {\bibfnamefont
  {P.}~\bibnamefont {Addesso}}, \bibinfo {author} {\bibfnamefont {R.~X.}\
  \bibnamefont {Adhikari}}, \bibinfo {author} {\bibfnamefont {V.~B.}\
  \bibnamefont {Adya}},  \emph {et~al.},\ }\href@noop {} {\bibfield  {journal}
  {\bibinfo  {journal} {Physical review letters}\ }\textbf {\bibinfo {volume}
  {121}},\ \bibinfo {pages} {161101} (\bibinfo {year} {2018})}\BibitemShut
  {NoStop}%
\bibitem [{\citenamefont {{Long}}\ \emph {et~al.}(2006)\citenamefont {{Long}},
  \citenamefont {{Van Giai}},\ and\ \citenamefont
  {{Meng}}}]{2006PhLB..640..150L}%
  \BibitemOpen
  \bibfield  {author} {\bibinfo {author} {\bibfnamefont {W.-H.}\ \bibnamefont
  {{Long}}}, \bibinfo {author} {\bibfnamefont {N.}~\bibnamefont {{Van Giai}}},
  \ and\ \bibinfo {author} {\bibfnamefont {J.}~\bibnamefont {{Meng}}},\ }\href
  {\doibase 10.1016/j.physletb.2006.07.064} {\bibfield  {journal} {\bibinfo
  {journal} {Physics Letters B}\ }\textbf {\bibinfo {volume} {640}},\ \bibinfo
  {pages} {150} (\bibinfo {year} {2006})},\ \Eprint
  {http://arxiv.org/abs/nucl-th/0512086} {arXiv:nucl-th/0512086 [nucl-th]}
  \BibitemShut {NoStop}%
\bibitem [{\citenamefont {Li}\ \emph {et~al.}(2015)\citenamefont {Li},
  \citenamefont {Margueron}, \citenamefont {Long},\ and\ \citenamefont
  {Van~Giai}}]{PhysRevC.92.014302}%
  \BibitemOpen
  \bibfield  {author} {\bibinfo {author} {\bibfnamefont {J.~J.}\ \bibnamefont
  {Li}}, \bibinfo {author} {\bibfnamefont {J.}~\bibnamefont {Margueron}},
  \bibinfo {author} {\bibfnamefont {W.~H.}\ \bibnamefont {Long}}, \ and\
  \bibinfo {author} {\bibfnamefont {N.}~\bibnamefont {Van~Giai}},\ }\href
  {\doibase 10.1103/PhysRevC.92.014302} {\bibfield  {journal} {\bibinfo
  {journal} {Phys. Rev. C}\ }\textbf {\bibinfo {volume} {92}},\ \bibinfo
  {pages} {014302} (\bibinfo {year} {2015})}\BibitemShut {NoStop}%
\bibitem [{\citenamefont {{Hofmann}}\ \emph {et~al.}(2001)\citenamefont
  {{Hofmann}}, \citenamefont {{Keil}},\ and\ \citenamefont
  {{Lenske}}}]{2001PhRvC..64b5804H}%
  \BibitemOpen
  \bibfield  {author} {\bibinfo {author} {\bibfnamefont {F.}~\bibnamefont
  {{Hofmann}}}, \bibinfo {author} {\bibfnamefont {C.~M.}\ \bibnamefont
  {{Keil}}}, \ and\ \bibinfo {author} {\bibfnamefont {H.}~\bibnamefont
  {{Lenske}}},\ }\href {\doibase 10.1103/PhysRevC.64.025804} {\bibfield
  {journal} {\bibinfo  {journal} {\prc}\ }\textbf {\bibinfo {volume} {64}},\
  \bibinfo {eid} {025804} (\bibinfo {year} {2001})},\ \Eprint
  {http://arxiv.org/abs/nucl-th/0008038} {arXiv:nucl-th/0008038 [nucl-th]}
  \BibitemShut {NoStop}%
\bibitem [{\citenamefont {{Baym}}\ \emph {et~al.}(1971)\citenamefont {{Baym}},
  \citenamefont {{Pethick}},\ and\ \citenamefont
  {{Sutherland}}}]{1971ApJ...170..299B}%
  \BibitemOpen
  \bibfield  {author} {\bibinfo {author} {\bibfnamefont {G.}~\bibnamefont
  {{Baym}}}, \bibinfo {author} {\bibfnamefont {C.}~\bibnamefont {{Pethick}}}, \
  and\ \bibinfo {author} {\bibfnamefont {P.}~\bibnamefont {{Sutherland}}},\
  }\href {\doibase 10.1086/151216} {\bibfield  {journal} {\bibinfo  {journal}
  {\apj}\ }\textbf {\bibinfo {volume} {170}},\ \bibinfo {pages} {299} (\bibinfo
  {year} {1971})}\BibitemShut {NoStop}%
\bibitem [{\citenamefont {{Negele}}\ and\ \citenamefont
  {{Vautherin}}(1973)}]{1973NuPhA.207..298N}%
  \BibitemOpen
  \bibfield  {author} {\bibinfo {author} {\bibfnamefont {J.~W.}\ \bibnamefont
  {{Negele}}}\ and\ \bibinfo {author} {\bibfnamefont {D.}~\bibnamefont
  {{Vautherin}}},\ }\href {\doibase 10.1016/0375-9474(73)90349-7} {\bibfield
  {journal} {\bibinfo  {journal} {\nphysa}\ }\textbf {\bibinfo {volume}
  {207}},\ \bibinfo {pages} {298} (\bibinfo {year} {1973})}\BibitemShut
  {NoStop}%
\bibitem [{\citenamefont {{Fortin}}\ \emph {et~al.}(2016)\citenamefont
  {{Fortin}}, \citenamefont {{Provid{\^e}ncia}}, \citenamefont {{Raduta}},
  \citenamefont {{Gulminelli}}, \citenamefont {{Zdunik}}, \citenamefont
  {{Haensel}},\ and\ \citenamefont {{Bejger}}}]{2016PhRvC..94c5804F}%
  \BibitemOpen
  \bibfield  {author} {\bibinfo {author} {\bibfnamefont {M.}~\bibnamefont
  {{Fortin}}}, \bibinfo {author} {\bibfnamefont {C.}~\bibnamefont
  {{Provid{\^e}ncia}}}, \bibinfo {author} {\bibfnamefont {A.~R.}\ \bibnamefont
  {{Raduta}}}, \bibinfo {author} {\bibfnamefont {F.}~\bibnamefont
  {{Gulminelli}}}, \bibinfo {author} {\bibfnamefont {J.~L.}\ \bibnamefont
  {{Zdunik}}}, \bibinfo {author} {\bibfnamefont {P.}~\bibnamefont {{Haensel}}},
  \ and\ \bibinfo {author} {\bibfnamefont {M.}~\bibnamefont {{Bejger}}},\
  }\href {\doibase 10.1103/PhysRevC.94.035804} {\bibfield  {journal} {\bibinfo
  {journal} {\prc}\ }\textbf {\bibinfo {volume} {94}},\ \bibinfo {eid} {035804}
  (\bibinfo {year} {2016})},\ \Eprint {http://arxiv.org/abs/1604.01944}
  {arXiv:1604.01944 [astro-ph.SR]} \BibitemShut {NoStop}%
\bibitem [{\citenamefont {Farhi}\ and\ \citenamefont
  {Jaffe}(1984)}]{PhysRevD.30.2379}%
  \BibitemOpen
  \bibfield  {author} {\bibinfo {author} {\bibfnamefont {E.}~\bibnamefont
  {Farhi}}\ and\ \bibinfo {author} {\bibfnamefont {R.~L.}\ \bibnamefont
  {Jaffe}},\ }\href {\doibase 10.1103/PhysRevD.30.2379} {\bibfield  {journal}
  {\bibinfo  {journal} {Phys. Rev. D}\ }\textbf {\bibinfo {volume} {30}},\
  \bibinfo {pages} {2379} (\bibinfo {year} {1984})}\BibitemShut {NoStop}%
\bibitem [{\citenamefont {{Kumar}}\ \emph {et~al.}(2022)\citenamefont
  {{Kumar}}, \citenamefont {{Thapa}},\ and\ \citenamefont
  {{Sinha}}}]{2022MNRAS.513.3788K}%
  \BibitemOpen
  \bibfield  {author} {\bibinfo {author} {\bibfnamefont {A.}~\bibnamefont
  {{Kumar}}}, \bibinfo {author} {\bibfnamefont {V.~B.}\ \bibnamefont
  {{Thapa}}}, \ and\ \bibinfo {author} {\bibfnamefont {M.}~\bibnamefont
  {{Sinha}}},\ }\href {\doibase 10.1093/mnras/stac1150} {\bibfield  {journal}
  {\bibinfo  {journal} {\mnras}\ }\textbf {\bibinfo {volume} {513}},\ \bibinfo
  {pages} {3788} (\bibinfo {year} {2022})},\ \Eprint
  {http://arxiv.org/abs/2204.11034} {arXiv:2204.11034 [astro-ph.HE]}
  \BibitemShut {NoStop}%
\bibitem [{\citenamefont {{Ju}}\ \emph {et~al.}(2021)\citenamefont {{Ju}},
  \citenamefont {{Hu}},\ and\ \citenamefont {{Shen}}}]{2021ApJ...923..250J}%
  \BibitemOpen
  \bibfield  {author} {\bibinfo {author} {\bibfnamefont {M.}~\bibnamefont
  {{Ju}}}, \bibinfo {author} {\bibfnamefont {J.}~\bibnamefont {{Hu}}}, \ and\
  \bibinfo {author} {\bibfnamefont {H.}~\bibnamefont {{Shen}}},\ }\href
  {\doibase 10.3847/1538-4357/ac30dd} {\bibfield  {journal} {\bibinfo
  {journal} {\apj}\ }\textbf {\bibinfo {volume} {923}},\ \bibinfo {eid} {250}
  (\bibinfo {year} {2021})},\ \Eprint {http://arxiv.org/abs/2111.08909}
  {arXiv:2111.08909 [nucl-th]} \BibitemShut {NoStop}%
\bibitem [{\citenamefont {{Lopes}}\ \emph
  {et~al.}(2021{\natexlab{b}})\citenamefont {{Lopes}}, \citenamefont
  {{Biesdorf}}, \citenamefont {{Marquez}},\ and\ \citenamefont
  {{Menezes}}}]{2021PhyS...96f5302L}%
  \BibitemOpen
  \bibfield  {author} {\bibinfo {author} {\bibfnamefont {L.~L.}\ \bibnamefont
  {{Lopes}}}, \bibinfo {author} {\bibfnamefont {C.}~\bibnamefont {{Biesdorf}}},
  \bibinfo {author} {\bibfnamefont {K.~D.}\ \bibnamefont {{Marquez}}}, \ and\
  \bibinfo {author} {\bibfnamefont {D.~P.}\ \bibnamefont {{Menezes}}},\ }\href
  {\doibase 10.1088/1402-4896/abef35} {\bibfield  {journal} {\bibinfo
  {journal} {\physscr}\ }\textbf {\bibinfo {volume} {96}},\ \bibinfo {eid}
  {065302} (\bibinfo {year} {2021}{\natexlab{b}})},\ \Eprint
  {http://arxiv.org/abs/2009.13552} {arXiv:2009.13552 [hep-ph]} \BibitemShut
  {NoStop}%
\bibitem [{\citenamefont {{Lopes}}\ \emph {et~al.}(2022)\citenamefont
  {{Lopes}}, \citenamefont {{Biesdorf}},\ and\ \citenamefont
  {{Menezes}}}]{2022MNRAS.512.5110L}%
  \BibitemOpen
  \bibfield  {author} {\bibinfo {author} {\bibfnamefont {L.~L.}\ \bibnamefont
  {{Lopes}}}, \bibinfo {author} {\bibfnamefont {C.}~\bibnamefont {{Biesdorf}}},
  \ and\ \bibinfo {author} {\bibfnamefont {D.~P.}\ \bibnamefont {{Menezes}}},\
  }\href {\doibase 10.1093/mnras/stac793} {\bibfield  {journal} {\bibinfo
  {journal} {\mnras}\ }\textbf {\bibinfo {volume} {512}},\ \bibinfo {pages}
  {5110} (\bibinfo {year} {2022})},\ \Eprint {http://arxiv.org/abs/2111.13732}
  {arXiv:2111.13732 [hep-ph]} \BibitemShut {NoStop}%
\bibitem [{\citenamefont {{Liu}}\ \emph {et~al.}(2001)\citenamefont {{Liu}},
  \citenamefont {{Gao}},\ and\ \citenamefont {{Guo}}}]{2001NuPhA.695..353L}%
  \BibitemOpen
  \bibfield  {author} {\bibinfo {author} {\bibfnamefont {Y.-x.}\ \bibnamefont
  {{Liu}}}, \bibinfo {author} {\bibfnamefont {D.-f.}\ \bibnamefont {{Gao}}}, \
  and\ \bibinfo {author} {\bibfnamefont {H.}~\bibnamefont {{Guo}}},\ }\href
  {\doibase 10.1016/S0375-9474(01)01120-4} {\bibfield  {journal} {\bibinfo
  {journal} {\nphysa}\ }\textbf {\bibinfo {volume} {695}},\ \bibinfo {pages}
  {353} (\bibinfo {year} {2001})},\ \Eprint
  {http://arxiv.org/abs/hep-ph/0105202} {arXiv:hep-ph/0105202 [hep-ph]}
  \BibitemShut {NoStop}%
\bibitem [{\citenamefont {{Liu}}\ \emph {et~al.}(2003)\citenamefont {{Liu}},
  \citenamefont {{Gao}}, \citenamefont {{Zhou}},\ and\ \citenamefont
  {{Guo}}}]{2003NuPhA.725..127L}%
  \BibitemOpen
  \bibfield  {author} {\bibinfo {author} {\bibfnamefont {Y.-x.}\ \bibnamefont
  {{Liu}}}, \bibinfo {author} {\bibfnamefont {D.-f.}\ \bibnamefont {{Gao}}},
  \bibinfo {author} {\bibfnamefont {J.-h.}\ \bibnamefont {{Zhou}}}, \ and\
  \bibinfo {author} {\bibfnamefont {H.}~\bibnamefont {{Guo}}},\ }\href
  {\doibase 10.1016/S0375-9474(03)01574-4} {\bibfield  {journal} {\bibinfo
  {journal} {\nphysa}\ }\textbf {\bibinfo {volume} {725}},\ \bibinfo {pages}
  {127} (\bibinfo {year} {2003})},\ \Eprint
  {http://arxiv.org/abs/nucl-th/0307058} {arXiv:nucl-th/0307058 [nucl-th]}
  \BibitemShut {NoStop}%
\bibitem [{\citenamefont {{Burgio}}\ \emph {et~al.}(2002)\citenamefont
  {{Burgio}}, \citenamefont {{Baldo}}, \citenamefont {{Sahu}},\ and\
  \citenamefont {{Schulze}}}]{2002PhRvC..66b5802B}%
  \BibitemOpen
  \bibfield  {author} {\bibinfo {author} {\bibfnamefont {G.~F.}\ \bibnamefont
  {{Burgio}}}, \bibinfo {author} {\bibfnamefont {M.}~\bibnamefont {{Baldo}}},
  \bibinfo {author} {\bibfnamefont {P.~K.}\ \bibnamefont {{Sahu}}}, \ and\
  \bibinfo {author} {\bibfnamefont {H.~J.}\ \bibnamefont {{Schulze}}},\ }\href
  {\doibase 10.1103/PhysRevC.66.025802} {\bibfield  {journal} {\bibinfo
  {journal} {\prc}\ }\textbf {\bibinfo {volume} {66}},\ \bibinfo {eid} {025802}
  (\bibinfo {year} {2002})},\ \Eprint {http://arxiv.org/abs/nucl-th/0206009}
  {arXiv:nucl-th/0206009 [nucl-th]} \BibitemShut {NoStop}%
\bibitem [{\citenamefont {{Yazdizadeh}}\ and\ \citenamefont
  {{Bordbar}}(2013)}]{2013Ap.....56..121Y}%
  \BibitemOpen
  \bibfield  {author} {\bibinfo {author} {\bibfnamefont {T.}~\bibnamefont
  {{Yazdizadeh}}}\ and\ \bibinfo {author} {\bibfnamefont {G.~H.}\ \bibnamefont
  {{Bordbar}}},\ }\href {\doibase 10.1007/s10511-013-9272-y} {\bibfield
  {journal} {\bibinfo  {journal} {Astrophysics}\ }\textbf {\bibinfo {volume}
  {56}},\ \bibinfo {pages} {121} (\bibinfo {year} {2013})},\ \Eprint
  {http://arxiv.org/abs/1303.1612} {arXiv:1303.1612 [astro-ph.SR]} \BibitemShut
  {NoStop}%
\bibitem [{\citenamefont {{Prasad}}\ and\ \citenamefont
  {{Bhalerao}}(2004)}]{2004PhRvD..69j3001P}%
  \BibitemOpen
  \bibfield  {author} {\bibinfo {author} {\bibfnamefont {N.}~\bibnamefont
  {{Prasad}}}\ and\ \bibinfo {author} {\bibfnamefont {R.~S.}\ \bibnamefont
  {{Bhalerao}}},\ }\href {\doibase 10.1103/PhysRevD.69.103001} {\bibfield
  {journal} {\bibinfo  {journal} {\prd}\ }\textbf {\bibinfo {volume} {69}},\
  \bibinfo {eid} {103001} (\bibinfo {year} {2004})},\ \Eprint
  {http://arxiv.org/abs/astro-ph/0309472} {arXiv:astro-ph/0309472 [astro-ph]}
  \BibitemShut {NoStop}%
\bibitem [{\citenamefont {{Contrera}}\ \emph {et~al.}(2022)\citenamefont
  {{Contrera}}, \citenamefont {{Blaschke}}, \citenamefont {{Carlomagno}},
  \citenamefont {{Grunfeld}},\ and\ \citenamefont
  {{Liebing}}}]{2022PhRvC.105d5808C}%
  \BibitemOpen
  \bibfield  {author} {\bibinfo {author} {\bibfnamefont {G.~A.}\ \bibnamefont
  {{Contrera}}}, \bibinfo {author} {\bibfnamefont {D.}~\bibnamefont
  {{Blaschke}}}, \bibinfo {author} {\bibfnamefont {J.~P.}\ \bibnamefont
  {{Carlomagno}}}, \bibinfo {author} {\bibfnamefont {A.~G.}\ \bibnamefont
  {{Grunfeld}}}, \ and\ \bibinfo {author} {\bibfnamefont {S.}~\bibnamefont
  {{Liebing}}},\ }\href {\doibase 10.1103/PhysRevC.105.045808} {\bibfield
  {journal} {\bibinfo  {journal} {\prc}\ }\textbf {\bibinfo {volume} {105}},\
  \bibinfo {eid} {045808} (\bibinfo {year} {2022})},\ \Eprint
  {http://arxiv.org/abs/2201.00477} {arXiv:2201.00477 [nucl-th]} \BibitemShut
  {NoStop}%
\bibitem [{\citenamefont {{Prasad}}\ and\ \citenamefont
  {{Mallick}}(2022)}]{2022MNRAS.516.1127P}%
  \BibitemOpen
  \bibfield  {author} {\bibinfo {author} {\bibfnamefont {R.}~\bibnamefont
  {{Prasad}}}\ and\ \bibinfo {author} {\bibfnamefont {R.}~\bibnamefont
  {{Mallick}}},\ }\href {\doibase 10.1093/mnras/stac2324} {\bibfield  {journal}
  {\bibinfo  {journal} {\mnras}\ }\textbf {\bibinfo {volume} {516}},\ \bibinfo
  {pages} {1127} (\bibinfo {year} {2022})},\ \Eprint
  {http://arxiv.org/abs/2207.03234} {arXiv:2207.03234 [astro-ph.HE]}
  \BibitemShut {NoStop}%
\end{thebibliography}%
\end{document}